*REVISTING THE CAPITALIST ROAD TO COMMUNISM: unconditional basic income and the post-labor world*

Robert van der Veen (University of Amsterdam) and Loek Groot (University of Utrecht)

ABSTRACT

The thesis of a *capitalist road to communism* (van der Veen and Van Parijs, 1986) asserts that Marx's realm of freedom can be reached from within welfare capitalism, skipping socialism, by using a tax-financed unconditional basic income until it is close to disposable income per head, so that the very distinction between paid work and free time is cancelled as a result. We revisit and update this thesis for two reasons: the recent prospect of a post-labor society following the automation revolution in technology, and that welfare capitalism has become more inegalitarian and less hospitable to basic income. We use a simple economic model which incorporates an upward adjustment of basic income to labor-saving technical change and distinguishes between capital that complements labor and capital that is fully substitutable with labor. A baseline simulation of the model shows the economic feasibility of a capitalist transition to communism. Two versions of a scenario incorporating interplay between technical change and market socialist institutional reforms are set out which make the transition politically viable to some extent, depending on the social distribution of power over technology. The most promising version is one in which the productivity of labor and automation capital grow at similar rates. We show in which respects it approximates the ideal of communism. One finding is that communism does not require reaching the final stage of a post-labor society. We conclude with a reflection on the relevance of our present update for the more immediate future of unconditional basic income.

**1. Introduction: why revisit the capitalist road to communism?**

In these interesting times we revisit an old idea, that Marx's communist ideal of material abundance, freedom from toil and distribution according to needs can be reached from within welfare capitalist economies. This requires harnessing the productivity and labor-saving potential of these economies by granting a steadily growing unconditional basic income to all citizens, and keep doing so, close to swallowing up disposable income per head and cancelling the distinction between paid work and free time. 'A Capitalist Road to Communism' was published in 1986 in a special issue of *Theory and Society*, commented on by six critics, and reprinted in *Basic Income Studies* in a special issue twenty years later in 2006, flanked again by the critical efforts of six other authors. Rather than trying to assess the merits of the original article - we call it *Capitalist Road* - and rather than engaging with the arguments of the twelve critics and the responses of the two authors, we shall discuss its main idea as it appears to us now, in 2024. In this introduction we explain three reasons for doing so, setting out how our story is organized along the way.

First, something like a 'capitalist road to communism' has become far more unlikely than it already seemed to be in 1986. The welfare state which would provide the framework for enabling this road to be traveled has been under severe pressure in the meantime. Closely related, inequality of wealth and income has grown in capitalist societies, with adverse consequences for political democracy, fair opportunities, social mobility, economic security, general well-being and human dignity. As a home for unconditional basic income to commence a transition to communism, welfare capitalism is no longer very hospitable, having become more Victorian in dispensing welfare and more capitalist in pursuit of private profit. Thomas Piketty's *Capital in the 21st Century* points to long-run trends underlying this - a rise in the share of income going to private capital, and the net rate of return to capital persistently



surpassing the growth rate of national income. These trends lead in the opposite direction of communism towards a plutocratic society with a modernized poor law.[1] We incorporate some of Piketty's points on the evolution of capitalism in our inquiry into the political feasibility of a transition to communism in sections 2.5 and 3.

Regarding the chances of installing a basic income at subsistence level, and going on to establish communism from there, the authors of *Capitalist Road* - duly prompted by several of their critics - realized that it would take large consensus on principles of social justice to give unconditional basic income the moral foundation needed for its institutional underpinning and lend democratic legitimacy to a policy of durably raising the grant to its economically sustainable level. Our present inquiry will largely bypass this major theme, which we addressed elsewhere in detail (van der Veen and Groot, 2019, 169-192).[2] We do however touch on the issue of social justice in sections 2.4 and 5.

Our second reason for wanting to revisit *Capitalist Road* is the promise of an increasing impact of automated processes, robotics and artificial intelligence - 'automation' for short - on the economy in general and on prospects for paid work and economic security.[3] In his materialist vision of how the capitalist mode of production would change, by way of a revolutionary shift to a future socialist mode of production enabling a transition to communism, Marx's primary assumption was that development of the productive forces in this final stage of history must inexorably lead to a falling demand for labor. As is well-known, Marx wrongly predicted in *Capital* that accumulation coupled to labor-saving productivity growth would forever continue to create a constantly growing reserve army of labor and produce massive immiseration of workers and unemployed amidst steeply rising income and wealth of capitalist owners (Marx, (n.d) (1887), ch. 25, sec. 4 and 5). Yet under contemporary forms of globalized capitalism including the welfare state variety, from which *Capitalist Road* took off in 1986, the forces of production might well develop rapidly enough to pose similar dangers.

---

[1] Standing (2023, Ch. 4-7) provides a minutely detailed account of how rentier capitalism and bureaucratic means-testing social security work together to produce forms of domination and uncertainty afflicting middle and poor workers, which amount to the appropriation of large portions of their time outside the wage contract.

[2] For the most recent formulation of how to defend basic income on grounds of justice see Van Parijs and Vanderborght 2017, Ch. 5. Their position is that setting unconditional basic income at its maximum sustainable level subject to rights of formal freedom is the only way to realize the largest possible set of income-leisure options for the least-advantaged in society. We share this liberal-egalitarian and Rawlsian position, while questioning the argument for the uniqueness argument for basic income on theoretical grounds. Finally, for reasons of focus and clarity, we adhere to the national scope of the original argument of *Capitalist Road*, and endorse the comments in the reply to authors (van der Veen and Van Parijs 2006, sec. 4) on the difficult problem of pursuing global justice through efforts to achieve justice on a national scale.

[3] See for example Suleyman, 2023, Ch. 6 and 7 on the powerful interaction between general-purpose technologies besides AI, such as quantum computing, synthetic biology and materials development, which leads to rapid technological change. Suleyman however does not affirm that these developments will necessarily lead to a world without labor, see 179-181.

It is true that technological unemployment today is most often believed to be just as unlikely as Marx was wrong in predicting a permanently disastrous effect of growth and labor-saving technological change after the industrial revolution in Great Britain. But an unprecedented wave of innovation fueled by automation is on its way today and expected to gain momentum, overturning the safe conviction that just as under previous major innovations, automation will create more new jobs than it replaces older ones. So Marx's primary assumption about economic growth involving a secular shortening of the 'labor day' could come true in coming decades, possibly to the point of ushering in an era of post-labor economics.

The provocative strategy of *Capitalist Road* was to take an advance on this last possibility, as a way of showing how an ideal of communism based on a freedom-oriented reading of the slogan *from each according to his ability, to each according to his needs* could conceivably be reached, without the need for replacing capitalism by a socialist regime of collective ownership run by the 'free association of producers'. Instead, the authors maintained this ideal



might be approximated (within an unspecified time interval) by starting from a welfare capitalist economy within a national state which grants unconditional income to all its citizens and residents. In the present twenties of the twenty-first century the prospect of entering a society with radically diminished need for paid labor has become more realistic. This makes us want to explore the idea of a capitalist road to communism in more depth. In section 2 we clarify the idea and relate it to normative interpretations of Marx's concepts of need-formation, abundance, alienation and exploitation, to show how a progressively larger share of basic income in national income might achieve the enrichment of human need on the basis of growing material opulence, the withering away of alienation in its different forms, and the elimination of the exploitation of 'man by man'.

In section 3 we present a stylized economic model. It incorporates basic income as a method of income redistribution, and specifies capitalist and market socialist ways of financing the basic income from taxes on (state-owned or private) capital and wages. The production function is adapted from the well-known Cobb-Douglas format. We call it the SA model since it was proposed by Lawrence Summers in 2013 and worked out by Anthony Atkinson in 2014.[4] The main feature is that automation capital is fully substitutable with labor, and both factors are complementary to fixed capital. In section 4, we use a simulation of this model to present an economically feasible time path, in which automation capital progressively replaces labor, as the Marxian transition to communism indeed requires.

The equations of the SA model can be specified by different modes of technological change which affect the productivity of labor and automation capital. In sections 4 and 5, we present two transition paths that correspond to these changes. As noted in section 2 and discussed in section 5 in more detail, the political feasibility of a capitalist transition to communism remains dubious, unless the institutions of welfare capitalism undergo social and economic reforms in order to counteract the inegalitarian tendencies of existing capitalism. Our view is that the transition to communism within one of several possible paths - including a path which involves strong productivity effects of automation on labor and a consequent slowdown of labor displacement - will not necessarily require the complete eradication of remunerated labor (section 5.4) nor a completely equal income distribution over time. To arrive at a communist society, it may be sufficient to get to the point where paid work (undertaken against small rewards) and free time (assisted by powerful automated devices) will still exist side by side, but have gained equal intrinsic value in many respects. This raises some questions about the relation between communism and a 'post-labor society' which we address at the end of section 5 before moving to our conclusions in the final section 6. We also have a third reason for revisiting *Capitalist Road*. How does our account of getting close to the ideal of communism in the not-too-distant future bear on the chances of getting an unconditional basic income implemented in the more immediate future? After having discussed our first two reasons, we reflect on this question in section 6.

---

[4] Summers, 2013, Atkinson, 2014, and 2015, 97, notes 32 and 33. The SA model as a tool for analyzing the conditions for the transition to communism was introduced in Van der Veen and Groot, 2019, 163-167.

## 2. The idea of *Capitalist Road* restated.

### 2.1 Reviving and revising Marx's view.

The idea of transition to a future stage of economic and social development - communism - in which workers are liberated from the compulsion to work for an alien (but human-powered) authority possessing the economic means of production and coercion through the wage relation - that idea comes from Marx's critique of capitalism, and from his commitment to the materialist interpretation of history. Marx was fascinated by the productive potential of capitalism and the capability of private owners to coerce, dominate and extract profit from



workers in a free contractual environment. He held that the development of productive forces would eventually be served by a revolutionary shift from capitalist market relations to a socialist economy, a shift which he thought to be historically inevitable. In his only vaguely outlined socialist economy, markets for production factors would be superseded by the authority of a spontaneously organized proletariat, the 'society of associated producers'. Collectively articulated personal needs would be satisfied partly by rationing and partly by allocating goods through limited consumer markets in keeping with the intention to overcome the remaining domination of workers by former members of the capitalist economic classes. Marx envisaged this form of socialist society as the stepping-stone towards a final stage of living together under communism. Socialism would pair economic growth to a shrinking need for labor, provide increasing levels of healthcare and education for all and 'funds for those who cannot work'. Socialist growth would satisfy increasingly enriched human needs by making available more and better means of consumption. It would bring continuous improvement of work quality in enterprises, with workers being more and more able to identify themselves with the process and the product of their own labor. This would amount to a continuously falling demand for labor conceived as *paid work*, ongoing de-alienation of laboring activity and - all within the same growth process - gradual replacement of the principle of labor contribution in income distribution by the principle of need. Ultimately, communism would join material abundance with freedom (Marx, 1977, 568-9).

To clarify in what sense the process leading to communism set out in *Capitalist Road* might be viewed as a desirable path from both a Marxian and a liberal-egalitarian standpoint, note that Marx's idea of the transition has been amended in three main ways. First, concerning property relations, the requirement of comprehensive planning on the basis of state ownership of land and capital could be dropped, following the discussions after Oskar Lange's argument



for casting collective ownership of the means of production in a competitive market environment. (Lange, 1938, 95-142). Forms of market socialism can thus be a possible starting point of a transition to communism.[5] Secondly, in the late nineteenth century, the revisionist social-democratic movement in Marxism led by Eduard Bernstein no longer wished to insist on state ownership of the means of production, relying on a gradual march of working class movements towards a socialist parliamentary system that could legally regulate and tax the capitalist economy. (Bernstein, 1909). Duly updated to match the development of the 20[th] century welfare state, a relatively egalitarian capitalist democracy with regulation of competition, labor law, and generous tax-financed provisions for health, education and housing could also be the transition's starting point.

In the normative sphere, Marx's own ideal of *self-realization through work*, ultimately to be attained from within socialism, has been qualified or even replaced by its companion ideal of *autonomous choice* (van der Veen, 1991, Ch. 2). This liberal conception of the transition to communism certainly does not dismiss the value of self-realization to be realized in the labor process. However, it no longer posits communism exclusively as a state in which 'labour has become not only a means of life but life's prime want' (Marx, 1977, 569). Instead, the ideal of autonomous choice will also recognize that projects and pastimes outside of paid (and indeed unpaid) work can be equally valuable, when conceived and executed in accordance with the reflectively self-chosen goals and interactions of individuals in society.[6] Given that the domain of paid labor is to become progressively smaller according to Marx, and income gets more decoupled from the performance of labor over time, this is a reasonable thought in the context of the transition to communism, though it is not one underwritten by orthodox Marxists or social democrats.

---

[5] See the different elaborations of market socialism in Bardhan and Roemer, 1993, Parts I and II. In Part III the authors include some progressive 'related alternatives to capitalism' which are compatible with market economies, such as labor-management of firms and associative democracy. In our list of market socialist reforms of capitalism in 5.1 below, we include these and similar alternatives.
[6] See van der Veen, 1991, Ch. 1.3 and Arneson, 1993, 292-295, in Bardhan and Roemer, 1993.

Thirdly, applied to market socialist or welfare capitalist regimes in a liberal-egalitarian conception of the communist ideal, the institutional device of an unconditional basic income has been put forward as a key instrument of reform for traveling the road to communism. This was the novel interpretation of the transition proposed in *Capitalist Road*. It is the one which we here develop further. Alternatively however, one might try to elaborate the final stage towards which a transition to communism might be headed from within either of the above two property regimes, while continuing to underwrite the orthodox view that self-realization through productive work within reciprocal relations of community is the sole good of humankind. As far as we are aware, this last kind of transition to communism has never been formulated. Its key instrument would presumably not be an unconditional basic income, but rather forms of a social wage and a job guarantee.

*2.2 The logic of the capitalist road and the role of basic income: preventing underemployment*

In the Marxian economy of time, the 'labor day' measures the proportion of time spent in paid work, or as Marx put it, in the 'realm of necessity'. Embedding basic income in the framework of a capitalist or market socialist welfare state provides a way of dealing with the risks and chances of an ever-shrinking realm of necessity. The risks are clear enough. When technological change seriously reduces the demand for paid labor, workers face involuntary underemployment, hence face being excluded from taking part in the labor market to a possibly large extent.[7]

---

[7] As Joan Robinson (1962/1974, 46) observed, "…the misery of being exploited by capitalists is nothing compared to the misery of not being exploited at all.".



Unconditional basic income will not only be an alternative source of economic security replacing the wage income of working families and propping up effective demand. If its level is calibrated adequately, basic income also offers the chance of preventing involuntary underemployment by providing a disincentive to work long hours, thereby adjusting the voluntary labor supply to match the ongoing decrease in the demand for labor at full capacity utilization. When labor demand keeps falling consistently over time due to technical change, this is a radical use of basic income. It implies that the level of basic income will have to grow year by year at a speed determined by how fast the demand for labor falls and how sensitive the average supplier of labor is to changes in the net wage rate and in the level of basic income. This use of basic income also implies that its level is the maximum sustainable one given the capital stock existing at each moment in time, and therefore functions as an instrument of redistribution that maximizes the income of the economically least advantaged members of society, i.e. those with no capital holdings and minimal wage-earning capacities. This can be taken to satisfy a version of the Rawlsian difference principle defined over income-leisure opportunities (see note 3 above). However, setting basic income to prevent involuntary underemployment is motivated independently, as we now explain.

Keeping the level of basic income to achieve equilibrium on the labor market at full capacity utilization is highly desirable from a Marxian perspective, for it contributes to reducing alienation in one important way within a welfare state with a democracy. As soon as technology is on course to replace labor and labor demand falls, basic income set at this optimal level enables workers to dispose of their own time. They are then on the way to having free exit from paid work, combined with free access to work.[8] This enables them to freely choose their preferred mixes of paid labor, care in the family, membership in self-chosen communities, participation in political organisations, and to engage in simple and complex leisure activities such as resting and playing.

---

[8] For an analysis of the relationship between basic income and involuntary underemployment which brings in a principle of compensatory justice, see Groot (2004, Ch. 2-3).

This particular enabling effect of basic income can be easily overlooked within a traditional Marxian perspective on alienation. That is because preventing involuntary underemployment does not address the most heavily emphasized aspects of alienation as a state in which workers are 'separated', or 'estranged' from their own activities of performing tasks within the labor process itself, and lack control over the products of their labor. But as we argue in 2.4 below, once basic income reaches a sufficiently high level, it can also address these well-known evils of capitalism. Eliminating involuntary underemployment by using basic income can reduce this major aspect of alienation - conceived above as lack of autonomous choice to navigate one's way in society - only gradually, for it will be hindered both by market and institutional failures. However, the goal fits well into Marx's more lofty understanding of communism, according to which individuals ultimately attain the many-sided capacities of the human 'species-being' in a social union of reciprocal need-satisfaction. As noted above, such lofty ideas are capable of being interpreted in a perfectionist fashion with heavy emphasis on the ethical good of productive work for oneself and others in communist society, or alternatively, in a more liberal way which stresses the 'thin' good for individuals under communism of acting autonomously in various domains after due reflection. The argument of *Capitalist Road*, which identifies unconditional basic income as a key instrument of the transition, favors the latter interpretation to a large extent, and so does our present restatement.[9]

---

[9] The contrast between these two interpretations of the good in a communist society is real, but its practical significance should not be exaggerated. The traditional perfectionist interpretation requires a large measure of real freedom of choice, while our liberal interpretation which stresses an alternatively perfectionist value of autonomy endorses the assignment of intrinsic worth to reflective choice and social awareness in making choices, and in that sense does not satisfy the liberal requirement of justificatory neutrality. Compare Kandiyali, 2022, with van der Veen, 1991, 1.3 and 6. Also see Susskind, 2023, who argues that justificatory neutrality with



Given that the demand for labor must fall during the transition to communism, the level of labor supply-reducing basic income in the capitalist welfare state must to rise over time. Moreover, basic income may also rise relative to wages and capital incomes, after the increased taxation needed to finance it has been deducted. It is by no means sure that this will happen, as is shown in sections 3 to 5 below, but when it does basic income links two of the main features of the transition to communism: the shortening of the labor day and the shift from contributions to needs in income distribution. We now comment on the latter feature.

*2.3 The logic of the capitalist road and the role of basic income: weak and strong abundance*

A transition to communism not only involves doing less paid labor and having more available free time. Qualitatively, the time spent outside of paid work will be supported by a rising supply of goods and services to satisfy an ever-expanding array of needs, needs which Marx thought to be 'many-sided and rich in nature'.[10] In a market socialist or capitalist setting, these enriched needs are being formed by the ways in which technological innovations interact with the provision of consumption goods within markets and mechanisms of public financing. Barring destructive world wars and ecological catastrophes (sad possibilities existing under any set of institutions from which we must abstract here), we follow the view of *Capitalist Road* that this process of market-led economic growth is capable of generating the formation and satisfaction of the enriched array of needs that Marx envisaged when saying that "…the productive forces have also increased with the all-round development of the individual, and all the springs of co-operative wealth flow more abundantly..." (Marx, 1977, 569).

---

respect to policy interventions in the field of AI which aim at giving work more meaning, and aim to make sense of meaning in a post-labor world, becomes more and more difficult to maintain for liberal thinkers.

[10] See in particular Heller, 1974, Ch. V, on the evolution of needs during the transition and their nature under communism.



As needs expand in general, the means of satisfying conventionally defined 'basic' needs of subsistence will also expand. Marx called these 'necessary needs', and tied them to the value of labor-power in capitalism.[11] He sometimes contrasted them to 'luxury needs', the objects of which are largely possessed by the ruling classes. But for the 'society of associated producers', he thought the relevant contrast to be between necessary needs and 'free needs', the objects of which satisfy more elevated human desires irrespective of class signaling, objects which are therefore to become available to all individuals as the forces of production develop. Those free needs arise as the result of interaction between the desires and the capacities of individuals in different social roles. They will be articulated to the extent that individuals learn to adapt to new technology and have access to 'free time devoted to many-sided activity' outside of the labor process.[12]

Our view on the role of basic income in a capitalist or market socialist transition to communism makes a major assumption not directly found in Marx's texts, but one which is in line with his ideas on the 'historical and moral' element in the concept of necessary needs. We assume that during the transition, per capita income and the customary norms of subsistence tend to grow in proportion. This entails that the per capita surplus above subsistence - a rough measure of material abundance with respect to the satisfaction of free needs - will also follow per capita income, hence that the material means of free needs will always be increasing in an absolute sense as long as per capita income grows.

---

[11] As is well known, Marx thought that necessary needs of subsistence underlying the value of labor power contain a 'historical and moral element' (Marx, (n.d.) (1887),167-8) which not only includes the goods needed to live a minimally decent life but also rudimentary means of educational and health care. Welfare capitalism provides these elements in large part in the public sector.
[12] See Heller, 1974, 33-36.



When Marx says that 'the springs of cooperative wealth flow more abundantly' in the 'higher phase of communist society', he does not say that communism does away with scarcity, or what amounts to the same, that human needs can be satisfied without alternative cost.[13] What Marx can reasonably be taken to say about abundance under communism is two things. First - and to repeat - that human needs have become 'many-sided and rich'. This can only mean that compared to past periods, those needs have been extended over more domains in life in more intricate and complex ways. To the extent that this happens, available wealth and income per person will surely become more plentiful than at present. But this does not yet differentiate communism from present society or from past societies. Secondly, taking off from the statement that what is 'flowing more abundantly' is the 'springs of *cooperative wealth*', abundance under communism should be construed as the availability of means for the satisfaction of free needs without requiring remunerated productive contributions from individuals in return. This certainly does differentiate communism from present and past societies.

Against this background, the capitalist road to communism requires that economic growth coupled to progressively less demand for labor must take a special form, such that, given the choices of free workers over time, the level of basic income may durably rise relative to per capita income. The share of basic income set at levels above subsistence will then increasingly be at the expense of the share going to net wages and net personal income from capital. In Marxian terms, the growth of the basic income share means that the principle of distribution according to need gradually becomes dominant over the principle of distribution according to (labor) contributions. This aspect of the transition process can be called the 'shift from contributions to needs' (van der Veen, 1984, 465-8; 1991, 107, 4.2 and 4.3).

---

[13] Abundance under communism cannot mean the end of scarcity. Turning resources of capital goods and technology into objects of need-satisfaction and using those objects to satisfy actual needs is subject to constraints of time and space.

Up to now we have talked loosely about the 'basic income share' as the extent to which 'each according to his needs' obtains during the transition. This is misleading if that share is equated to the share of unconditional cash payments in annual income. For to have its liberating and enabling effects, an unconditional basic income must be embedded in the public provisions of an extended welfare state organized along egalitarian lines. In our treatment of the transition to communism in the economic model discussed below, 'distribution according to needs' therefore involves a system of *dual provision* (Van der Veen and Groot, 2019, 159-160). Individual welfare claims to income in both cash and kind of various types ranging from health, education and culture to care for the disabled must have a prominent place alongside entitlements to cash income which are dispensed universally and unconditionally to all citizens and residents. This requirement of dual provision especially applies to coverage of necessary needs, the means of which are expected to grow in tandem with per capita income. Necessary needs, then, are to be satisfied by variable welfare dispensations based on rules of equal access coupled with periodically paid unconditional cash transfers.

The distinction between weak and strong abundance made in *Capitalist Road* can now be clarified as follows. Weak abundance is a state of the economy in which necessary needs are just covered unconditionally. The share of basic income and welfare dispensations then exactly covers the ruling norm of subsistence which defines necessary needs and is assumed to be a constant fraction of per capita income according to our interpretation of what Marx regarded as the historical and moral element of the value of labor power. Whether this norm governing subsistence income is explicitly recognized, and if so, what institutional form it



takes does not need to be discussed at this point. What is important is the salience of weak abundance, the reasons why this state of the economy is singled out as the starting point of the transition to communism. Conceptually, it is easy to see this: the 'weak' stage of abundance begins when the principle of need unconditionally covers necessary needs only, whereas the 'strong' stage is entered when all needs are covered by the same principle, and when additionally, the very distinction between necessary and free needs has become irrelevant.

*2.4 Weak abundance, alienation and exploitation*

There is also an important substantive reason to focus on weak abundance: its beneficial effects on workers. From the point onwards at which subsistence is granted unconditionally, the twin ills of capitalism - alienation and exploitation of workers - will be in the process of being reduced or even eliminated. Regarding alienated labor, this is explained as follows. Once the level of basic income generating equilibrium on the labor market rises and starts to provide subsistence, individual workers have a credible option to leave work which does not suit their needs and preferences. The credibility of this option may or may not be supported by labor unions. But in any case the bargaining power of workers conferred by a subsistence-covering basic income will be improved, both in individual and collective negotiations with firms and managers in governmental organizations. This will tend to raise the quality of paid work obtained by the workers from their employers and/or speed up the replacement of low-quality jobs. As a result of these interactions within the workplace, both in drawing up the labor contract and subsequently performing contacted labor, workers will be better able to identify with the work processes and the final products in which they freely choose to be involved. This means that they will tend to lead a less alienated existence in the realm of necessity. When referring loosely to a measured 'index of work quality' in 3.1 below, we are thinking about the following three senses of labor quality inspired by Marx's alienation



concept. The dimensions of such an index are: control over your own labor process in collaboration with fellow workers, to be well-informed about the product of your labor and endorse it as meaningful, and the knowledge that your conditions of laboring have been, and are being, 'worked out' on a fair basis.[14]

Weak abundance is also of importance for ending or at least reducing the exploitation of workers. Defined by Marx as the 'extraction of surplus labor', exploitation can be understood in two distinct senses. First, if the focus is on the *extraction* of surplus labor from workers, interpreted as situations in which they are economically or politically compelled to enter a wage contract - a contract which inevitably involves performing surplus labor - then the existence of a credible option to subsist without having to do any paid work at all eliminates exploitation by definition, because the surplus labor relinquished by the worker, in the absence of the relevant compulsion, is not 'extracted'. Once the economy becomes weakly abundant, then, exploitation of workers in this freedom-oriented sense has ceased as the transition to communism begins. Secondly, if the focus is on the value of the *surplus labor* which the employer appropriates from the worker, and exploitation could be said to occur even when workers voluntarily choose to work, it then refers to the degree to which the surplus value appropriated by the employer fails to benefit the worker, directly or indirectly. In this second sense of Marxian exploitation, bargaining power conferred by credible options to leave the labor contract will then promote arrangements of sharing the benefits of surplus value in more or less equitable ways among workers and employers. Reaching weak abundance will therefore tend to reduce the exploitation of workers in this distributive sense, and for that reason also may be a natural starting point for the transition to communism.[15] Of course, the extent to which weak abundance enables a worker to share the benefits of surplus value in a particular contract depends on many background factors - the power of organized labor being a major one.

---

[14] Labor quality so conceived is a 'midstream' notion, capable of being concretely specified in different ways. Going into it upstream are abstract constitutive notions of value such as purpose and meaning, coming out of it downstream are consequential features of the three dimensions above, for example whether the work is (sufficiently) challenging, not too repetitive and stressful, occurring in pleasurable surroundings, providing recognition, etc. This definition of *quality* is not exclusively reserved for labor. One could simply replace 'labor' with some other non-labor category of activity, for example the quality (for a person) of 'unpaid care', 'running a family', or 'taking part in collective desisionmaking'. This should be kept in mind when discussing moments in future times located closer to communism, in which paid work is strongly reduced, see 5.3 and 5.4 below.

Summarized in operational terms, the transition to communism involves economic and social development along four dimensions, growth of per capita national income, decrease of the demand for labor (matched by decrease in the supply of labor), increase in the quality of paid labor, and increase in the share of social expenditure and unconditional basic income in per capita income. The transition proper can be said to start at the point of weak abundance, when workers have the credible option to do without a labor contract because their 'necessary needs' of subsistence are covered by basic income (please note that 'basic income' will be our often-used shorthand for the mix of egalitarian social outlays paid in cash and kind, augmented by transfers of unconditional cash income for all).

Ultimately, development along the four dimensions of the transition culminates in something that might be called *full communism*, the phase of society in which distribution according to needs has become dominant and strong abundance therefore obtains, and also a phase in which production virtually occurs without the incentives of payments and perks. Finally and perhaps most importantly, full communism refers to a society in which the very distinction between activities of production and consumption has become blurred and in most respects irrelevant. Developments along the four dimensions of the transition go together to a large extent. But it does not need to be the case, for example, that basic income must take up the whole of per capita income at the same moment at which the demand for labor has dropped to zero. Moreover, it is not even a requirement of full communism that 'working for money' has been eliminated altogether. What would be required instead is that the work will be done just as much for its own sake as for the money, even if it might be true in some cases that the work would not be done, or less of it would be done, without the promise of money. In section 5 of this article we try to illustrate these points in some detail.

---

[15] van der Veen, (1991, Ch. 5, 157-9) argues for the first sense of Marxian exploitation, but notes that in the context of capitalism, Marx used a canonical definition of *class exploitation*. Workers depend on their wage only. Subject to the threat of destitution on being fired, they are compelled to perform surplus labor for the capitalists, the entire product of which is appropriated by the capitalists as investment and personal consumption (see 157-159). Class exploitation therefore joins the freedom-oriented and distributive senses of the concept.

Marx talked about communism as a 'higher phase' of society rather than a static endpoint, and was understandably vague about the timescale of the transition to this higher phase. In the context of present welfare capitalist societies, that timescale matters much more than it did for Marx, especially if it turns out that rapid technological change remains as prominent as it is in the first decades of the twenty-first century. In fact, issues concerning the time in which full communism might actually be materializing along the above four dimensions were also largely absent from the article we are revisiting here, as evidenced by the statement in *Capitalist Road* that "…communism only constitutes the notional and unreachable end of a transition that is bound to drag on forever, even under the most favorable circumstances." (van der Veen and Van Parijs 1986a, 641).

*2.5 The political viability of the capitalist road and the relevance of market socialism*

Sections 3 and 4 below model the capitalist road to communism using a simulation method, provide an analysis of how the four dimensions of the transition hang together, and demonstrate the economic feasibility of traveling along the road. But as noted above, there are reasons to doubt the political viability of a social development turning around an



unconditional basic income which expressly aims to take up an ever-larger share of national income, even apart from obvious constraints posed by the need to steer a transition in the direction of climate and ecological safety and military security. The transition to communism may also be postponed indefinitely if the point of weak abundance is thought to be far away in time. This may happen when the tax capacity of welfare states is limited, and the new and controversial instrument of unconditional basic income is perceived to be in competition with long-existing legitimate claims of welfare state provisions which also demand increasing attention in an aging population. Moreover, the level of unconditional transfers compatible with keeping the labor supply up to full-capacity requirements when technical change is not labor-saving may be insufficient to cover subsistence and permanently need to be complemented by conditional transfers of social security, thereby undermining the rationale of adding an unconditional basic income to the mix of welfare-state arrangements (van der Veen and Groot, 2019, 159-163). The original aim of *Capitalist Road*, which was devoted to setting out the novel idea of a basic income-supported transition to communism understandably bypassed such complicating factors.

However, what most endangers a transition path based on a continuous rise of basic income is large and persistent economic inequality, our first reason for revisiting *Capitalist Road*. In *Capital in the Twenty-First Century* (2014), Thomas Piketty identified two long-term 'laws' or tendencies of capitalism: the rise of the share of capital in total income and the tendency of the rate of economic growth to lie below the net rate of return to capital. Among other important factors, notably the existence of an internationally mobile layer of highly remunerated workers, these two tendencies contribute to inequality of wealth and income when personal wealth and income holdings are distributed unequally to begin with. The dangers of these tendencies are also highlighted in a world-historical study by Bas van Bavel.



In *The Invisible Hand?* (2016), van Bavel documents a cyclical pattern in market economies which goes through history from early medieval Iraq onwards, from a spontaneous rise of factor markets on the basis of a relatively open and participatory state of society to intensive accumulation of capital, followed by growing inequality of personal wealth and capital, monopolization of firms and financialization of investment, and a final stage in which political capture of the state by large property owners goes together with institutional sclerosis. After this, decline of the market system sets in. Van Bavel extends this succession of cycles to England and the United States between 1800 and the present.

Our response to the deterministic narratives of Piketty's *Capital* and Van Bavel's *The Invisible Hand?* is voluntaristic. We will follow the approach taken by Piketty himself in 2019, in *Capital and Ideology,* his one-thousand-page historical overview of social justifications of inequality. In the final chapter on a possible future evolution towards more equality, Piketty outlines an engaging set of 'elements for a participatory socialism'. We use some of these elements to identify a stylized market socialism in sections 3 and 5 below which serves as a pointer for reforming the welfare capitalism sketched in *Capitalist Road*, assuming that a basic income is accepted, and weak abundance is within (not too distant) reach. But this is not an attempt to revive a market socialist road to communism. As the transition proceeds, it may not matter much anymore whether we talk about market socialism or about a capitalist system which is embedded within egalitarian institutions to support the key policy of basic income.

## 3. The Summers-Atkinson model of growth and technical change

*3.1 Presentation of the model*

The Summers-Atkinson (abbreviated SA) model of growth and technical change introduced in 1.1 is presented in Textbox 1 for the capitalist and market socialist formats.



Textbox 1: The SA model for welfare capitalism (*CAP*) and market socialism (*MS*)

| | |
|---|---|
| **CAP model** | **MS model** |

$$(1)\ Y = K_1^\alpha\ (AL + BK_2)^{1-\alpha}$$

$$(2)\ \alpha Y = rK_1$$

$$(3)\ (1-\alpha)AY = w(AL + BK_2)$$

$$(4)\ (1-\alpha)BY = r(AL + BK_2)$$

$$(5)\ nG = t(rK + wL) \qquad\qquad nG = rK\ +\ t_w wL \qquad (5')$$

$$(6)\ L = n - \delta G/w(1-t)Q \qquad\qquad L = n - \delta G/w(1 - t_w)Q \qquad (6')$$

$$(7)\ I = sY$$

$$(8)\ K = K_1 + K_2$$

$$(9)\ Y = wL + rK$$

$$(10)\ K_{T+1} = (1-\theta)K_T + I_T$$

In the production function of equation (1), total value of output $Y$ is the result of combining capital and labor. The capital stock consists of two types, $K_1$ and $K_2$. Summing the two gives the total capital stock $K$, as in equation (8). $K_1$ represents fixed capital goods such as buildings and machinery. Capital goods of type $K_2$ represent automated processes which are able to perform tasks that humans have hitherto performed with at least equal efficacy. $K_2$ is assumed to be fully substitutable with labor $L$. One unit of $K_2$ is equivalent to $B/A$ units of labor, where the technical parameters $A$ and $B$ respectively denote the productivity of labor and of capital of type 2 at a given moment in time. $K_1$ is complementary to the mix of labor and automation capital $(AL + BK_2)$ currently in use. In the production function the degree of complementarity in producing a given amount of output value $Y$ is expressed by parameters $\alpha$ and $(1 - \alpha)$ in the exponents of $K_1$ and $(AL + BK_2)$ respectively.

Differentiating the residue of benefits over costs $Y - wL - r(K_1 + K_2)$ to $K_1, L$ and $K_2$ gives three first order conditions in equations (2), (3) and (4), which determine the efficient



allocation of inputs, with the returns to capital $r$ and labor $w$ when the system is in general equilibrium. Given a rental price of capital, $r$, firms are indifferent to invest in $K_2$ or hire labor if $w/r = A/B$, where $w$ is the wage per unit of labor. At the margin, a unit of labor contributes $A$ against a cost of $w$, while an extra unit of automation capital $K_2$ generates $B$ against a cost of $r$. As long as $w/r < A/B$ , firms will not invest in automation, but as soon equality obtains investment in $K_2$ will start and $w/r = A/B$ will be maintained. We assume in section 4 that $A$ and $B$ change at a constant rate reflecting the progress of technology.

Total income $Y$ consists of gross wage income $wL$ and capital income $rK$, as in equation (9). In the capitalist format of the SA model ($CAP$), $Y$ is taxed at a uniform rate $t$ to finance social expenses including the unconditional basic income. Taken together, these expenses are denoted by $nG$, as in equation (5), where $n$ is the population and $G$ the sum of welfare expenditures and basic income. The textbox also identifies a market socialist format of the SA model ($MS$), which finances part of $nG$ from the income $rK$, assuming that $K$ represents state-owned capital which is rented out to autonomous firms. The remaining part is financed by a uniform tax $t_w$ on wage income $w$, as in equation (5').

The labor supply function for the $CAP$ and $MS$ formats, shown respectively in equations (6) and (6'), is a linear expression consisting of the maximum number of hours per period of a standard person (normalized at 1 unit) times the constant population $n$, minus a disincentive term.[16] In the $CAP$ format, this disincentive term $\delta G/w(1-t)Q$ contains the ratio of social expenses and basic income $G$ to the after-tax wage rate $w(1-t)$, which gets weighted by a term $Q$ expressing the quality of labor. The ratio is then multiplied by a parameter $\delta$ which denotes the propensity of the average worker to changes in $G$, $w$ and $Q$. Thus $\delta$ measures how sensitive the standard worker's willingness to supply labor is to the ratio of basic income and the quality-weighted net wage rate in each of the two formats. In section 4 we assume for simplicity that $Q$ changes over time at a constant rate. Given that the *MS* format taxes wage income at rate $t_w$, the disincentive term of (6') equals $\delta G / w(1 - t_w)Q$ .

---

[16] Atkinson (1995, 44) uses a labor supply function of the same form.

In both *CAP* and *MS*, investment is from savings out of income. The rate of saving *s* is defined as a constant fraction of total income *Y,* as in equation (7).[17] Finally, the intertemporal equation (10) affixes a time index (*T*). It says that the capital stock of the next period $K_{T+1}$ equals the capital of the current period $K_T$ minus a constant depreciation factor $\theta K_T$ , to which the investment of the current period $sY_T$ is added, as equation (7) prescribes. In section 4 we show how the equation systems of these two institutional formats behave over time. The equations of the two model versions, *CAP* and *MS*, differ only with respect to the disincentive terms of labor supply, given their different ways of financing the social expenses and unconditional basic income *G*. Underlying this are major institutional differences which bear on rules of public ownership and on the expected distributions of income and personal wealth of these two model formats. As explained in 2.5 above, the (presently non-existent) *MS* format of the SA model has been added as a marker pointing to possible ways of reforming 'really existing' welfare capitalism to increase the likelihood of a transition to communism. In section 5 we go into this in detail.

To prepare further discussion the following details may be helpful. First, *MS* and *CAP* have ten equations and ten endogenous variables $Y, K_1, K_2, K, r, w, L, G, I$, where *CAP* has the endogenous tax rate *t* and MS has tax rate $t_w$. We fix a value for *K* in *T* = 0 and set $K_2 = 0$.

---





This enables solving for $K_{t+1}$ in $T = 1$. The system of equations in Textbox 1 can be simulated numerically. To show that capital of type 2 is complementary with capital of type 1 and fully substitutable with labor, note that in equilibrium, the marginal products of capital are equal i.e. ( $\frac{\partial Y}{\partial K_1} = \frac{\partial Y}{\partial K_2}$ ). This gives $K_1 = \frac{\alpha[AL + BK_2]}{(1-\alpha)B}$ , or equivalently $K_2 + \frac{A}{B}L = \frac{(1-\alpha)}{\alpha}K_1$. Among the unknown variables, $w$ and $r$ are the only ones that can be expressed in 'reduced form', that is, solely in terms of the technical parameters. Dividing (3) and (4) gives $\frac{w}{r} = A/B$, which inserted in (2) gives $K_1 = \alpha AY/Bw$. Substituting this in (1) gives $Y^{1-\alpha} = \left(\frac{\alpha A}{Bw}\right)^\alpha (AL + BK_2)^{1-\alpha}$. Rewriting (3) as $Y = \frac{w(AL + BK_2)}{(1-\alpha)A}$ and inserting in the previous expression gives $w = A\alpha^\alpha(1-\alpha)^{1-\alpha}/B^\alpha$. From $\frac{w}{r} = A/B$, $r = B^{1-\alpha}\alpha^\alpha(1-\alpha)^{1-\alpha}$.

Finally, the capitalist or market socialist road to communism is operationalized in six criteria. The first four come from our exposition of the transition in 2.4. They are headed '$M$' to show their Marxian origin:

$M1$: total income $Y$ rises

$M2$: labor input $L$ falls

$M3$: the index for quality of labor $Q$ rises

$M4$: the share $G$ of basic income and social expenditure in total output $nG/Y$ rises.

The last two transition criteria, headed by '$P$' were introduced in 2.5. They are imposed by Piketty's account of the long-term tendencies of capitalism, which together contribute to mounting personal inequality of wealth and income.

$P5$ : the capital share of income $rK/Y$ rises

$P6$ : at all times, the net rate of return on capital $r(1 - t)$ lies above the growth rate $g$ of output.



These two criteria properly belong to the *CAP* version of the model and have a different meaning in the (non-existent but doctrinally relevant) market socialist MS model.

*3.2 Contrast with other models dealing with the impact of automation on work*

To study technological progress, the canonical model is an aggregate Cobb-Douglas or CES production function with capital and labor as factor inputs. In such models, capital complements labor by construction. As Summers (2013, 4) remarks, "Output is a function of capital and labor. Capital augments labor: it raises the productivity of labor. If there are only two factors, they have to be complements." Acemoglu and Restrepo (2018a) show that if technological progress is capital-augmenting then, following the consensus in empirical findings of a moderate substitution elasticity between capital and labor, it increases the wage rate as well as labor demand and the labor share.[18] This however is at odds with recent empirical evidence on the effects of automation on labor.

Two approaches can be distinguished to accommodate the mounting empirical evidence on automation, which besides a declining aggregate labor share also includes technological change in favor of the high-skilled[19], declining absolute wages for the low skilled[20] and job polarization.[21] The first is the task-based approach which we discuss below. The other approach, which we follow in 3.1, is to modify the aggregate production function along the lines suggested by Summers (2013) and Atkinson (2014; 2015).[22] Atkinson, the pioneer who took up Summers' suggestion to make a part of $K$ fully substitutable with $L$ in his 2014 article later expressed his trust in the capacity of this model to explain the rise of the capital share independently of the elasticity of substitution (Atkinson, 2015, 97-98).

---

[18] Under capital-augmenting technical change, the productivity coefficient $A_k$ of capital $K$ increases in the production function $Y = F(A_k K, A_l L)$.

[19] For example, the widespread introduction of the personal computer on the work floor raised the demand and wages for high-skilled while outweighing the increasing supply of skilled labor.

[20] Susskind (2022, 644) refers to Autor (2019) for evidence on declining wages of non-college workers.

[21] See Goos, Manning and Salomons (2014).

[22] The SA model is not mentioned much in the literature on AI and work. We have found only one case of a closely similar model in Korinek and Stiglitz (2019, Box 14.1, 372).

The task-based approach is a powerful microeconomic device for tracing the different channels through which automation might shape the future of work. In this approach, production factors are inputs to perform tasks, and tasks then combine to produce goods. In analyzing the ways in which automation proceeds, researchers initially distinguished routine from more intricate non-routine tasks, claiming that routine tasks are far more susceptible to replace labor by automation capital. This originally helped to explain the phenomenon of polarization, where the wages and share of medium-skilled workers declined, in part because their jobs contained many routine tasks that could be easily automated. More recently, the claim that non-routine tasks are immune to automation is under pressure due to the emergence of AI systems based on 'large language models' which are trained on huge datasets and capable of performing operations thought to be restricted to humans. In the newer task-based applications, therefore, there is a shifting cut-off demarcating tasks that can and cannot be automated instead of a fixed dichotomy.

Although the automation of previously non-automated tasks can intensify labor-saving in the economy, Acemoglu and Restrepo (2018b) identify important countervailing economic effects. First, due to lower costs, the labor demand for non-automated tasks rises. Secondly, with more capital accumulation, the existing capital stock becomes more productive. Thirdly, new tasks emerge, in which it is commonly assumed that labour has a comparative advantage (see also Susskind 2020; 2021). However, as Korinek discusses (2023, 32) this last effect may disappear in a future of work in which AI eventually reaches artificial general intelligence (AGI) by surpassing all human capabilities, so that labor loses its comparative advantage



altogether.[23] According to Korinek, scenarios in which AGI is reached within 20 or even 5 years from now should be taken seriously by policymakers. Recently, Acemoglu (2024, sec. 3) questions such scenarios and expects only very small total factor productivity effects of automation in the coming decade.[24] But even if the advent of AGI occurs in a more distant future, it is still advisable to think about the role basic income might play in protecting the livelihood of displaced workers. At several places, Korinek and co-authors argue that the more labour becomes redundant, the more reason there will be for providing unconditional income support to workers. They propose a so called 'seed UBI' which can be scaled up in case the capital share in the economy increases due to automation.[25]

We acknowledge the power of the task-based approach to handle the many different issues involved in assessing the impact of automation on the future of work. However, for our purposes of analyzing the role of basic income in the context of the capitalist road to communism - a road which requires a declining labor demand and an increasing capital share according to our transition criteria *M2* and *P5* - it is important to use a simple model with an aggregate production function, which enables us to distinguish in principle the effects of different ways in which automation affects the economy. The SA model is fit for this purpose.

## 4. The transition to communism in the SA model

### 4.1 Economic feasibility of the transition: a baseline simulation

We simulate the SA model of section 3.1 for a capitalist or market socialist economy which is on the verge of implementing $K_2$ by introducing automation in the production process. This happens in period $T = 1$. In the initial period $T = 0$ only $K_1$ is used. Using arguably reasonable initial values and parameters, we investigate whether the six transition criteria identified in 3.1 are satisfied. Recall that criterion *M1* prescribes continuously rising $Y$. In capitalism, this growth goes together with *P5,* which says that the capital share $rK/Y$ must increase. Moreover, according to *P6,* the net reward per unit of capital should always exceed the growth rate $g$ of national income, i.e. $r(1 - t) > g$.[26] During the transition, $L$ should fall over time and the ratio $nG/Y$ must rise. These last two criteria are imposed, respectively, by *M2*, the 'shortening of the working day' and *M4*, the 'shift from contributions to needs'.

---

[23] See Korinek 2023: 32, chart 2 which assumes finite brain power and an advancing frontier of automation.
[24] Acemoglu, D. (2024), The Simple Macroeconomics of AI. MIT, Working Papers in Economics, May. Retrieved 16-6-2024.
[25] See Korinek & Juelfs 2022, 4, 28-29, 33-34, Korinek 2023, 4, Korinek & Stiglitz 2019, 371-2, Klinova & Korinek 2021, 646.

Regarding criterion *M3*, we have simplified our story about the rising quality of labor in the subsection on alienation in 2.4. There we argued how increases in the level of basic income, when set to equate the supply and demand for labor under weak abundance, tend to raise the quality of labor in firms and public organisations by improving the bargaining power of workers. Instead of specifying the causal mechanisms of this tendency in separate equations which would incorporate a labor quality index $Q$ as an unknown of the model, we assumed that $Q$ is a parameter which increases at a constant rate during the transition. Obviously, this is too crude. Moreover, $Q$ is not only determined by the bargaining power of workers but also by the nature of the changing technology itself, as we will see in section 5.[27]

The initial values and parameters in the baseline scenario of Table 1 below are as follows:

- In $T = 0$, $K = 500$ and $L = 75$.

---

[26] In addition to the two Piketty transition criteria, another 'law' mentioned in *Capital* bears on the steady state property of the SA model, namely that $K/Y$ approaches $s /(g + \theta)$ . See Piketty, 2014, Ch. 5.

[27] In view of these limitations labor quality might be omitted from the model. We choose not to do so because of the importance of labor quality in Marxian communism, however conceived. When omitting $Q$ it must be shown that our simulations satisfy the other transition criteria. We checked that this is the case for the initial values and parameters shown in Tables 1 to 3. The difference is that the levels of $G$ and $t$ will be somewhat lower.



- The exponent of the production function is $\alpha = 0,25$. In $T = 0$, the parameter $\alpha$ equals the capital-income share $rK/Y$ from equations (2) and (8). In that initial period, $Y$ is given by a standard Cobb-Douglas production function with exponents 0,25 for $K$ and 0,75 for $L$.

- In the initial period, we set $Y$ to 100. Given $\alpha$ and the initial value for $K$, and $K_2$ still at zero, it follows that the initial value of labor productivity in (1) is $A = 0,779$. [28] In the baseline scenario shown below, we assume parameter $A$ changes at a rate $g_A = 0,01$ or 1 percent per year. From equation (2), it follows that in $T = 0$ the return to capital $r$ equals 0,05 or 5 percent.

- From equation (4) and given that initially $K_2 = 0$, the productivity parameter of automation capital is $B = 0,039$. We assume $g_B = 0,015$ per year.

- Quality of labor index $Q$: 1,00 in $T = 0$, with rate of change $g_Q = 0,005$.

- Population: $n = 100$. Savings rate out of $Y$: $s = 0,15$.[29]

- Depreciation rate of capital goods: $\theta = 0,02$.

- Sensitivity parameter of labor supply: $\delta = 25$, which corresponds to an initial labor supply elasticity of 1/3.

The simulation proceeds as follows. Given the initial value of $K = 500$ and the values of the parameters shown above, the values of the unknowns are selected by varying $Y$ in each period so that the equations of Textbox 1 are satisfied. Table 1 shows the values for a run up to 100 periods, calculated in Excel. Period $T = 0$ has the same values for $MS$ and $CAP$, except for $G$, which is at all times smaller in $CAP$ than in $MS$ because the net wage rate of $CAP$ depends on $t$ in equation (6) rather than on the smaller $t_w$ as in (6').

---

[28] With $K = 500$ and $Y = 100$, this corresponds to a capital-income ratio of 5, which is in line with an advanced welfare state such as the Netherlands in 2019, see https://wid.world/document/household-wealth-and-its-distribution-in-the-netherlands-1854-2019-world-inequality-lab-working-paper-2022-19/.

[29] Empirically, the savings rate varies. For example in the Netherlands it lies between 10 and 25% between 1995 and 2020, see https://opendata.cbs.nl/statline/#/CBS/nl/dataset/84097NED/table?ts=1703163023509.



Table 1. Simulation of the transition to communism in a baseline scenario

| T | Y | $K_1$ | $K_2$ | K | L | r | w | $G_{CAP}$ | t | $G_{MS}$ | $t_w$ |
|---|---|---|---|---|---|---|---|---|---|---|---|
| 0 | 100 | 500 | 0 | 500 | 75 | 5,00% | 1,00 | 0,50 | 0,50 | 0,57 | 0,43 |
| 10 | 108 | 485 | 66 | 551 | 73 | 5,59% | 1,06 | 0,57 | 0,53 | 0,66 | 0,45 |
| 20 | 118 | 473 | 132 | 605 | 71 | 6,25% | 1,13 | 0,65 | 0,55 | 0,76 | 0,48 |
| 30 | 130 | 463 | 200 | 664 | 69 | 6,99% | 1,21 | 0,74 | 0,57 | 0,88 | 0,50 |
| 40 | 142 | 456 | 272 | 728 | 67 | 7,82% | 1,28 | 0,85 | 0,59 | 1,01 | 0,52 |
| 50 | 157 | 449 | 350 | 799 | 64 | 8,74% | 1,37 | 0,97 | 0,62 | 1,17 | 0,54 |
| 60 | 174 | 444 | 435 | 879 | 60 | 9,77% | 1,45 | 1,11 | 0,64 | 1,35 | 0,56 |
| 70 | 193 | 441 | 528 | 968 | 56 | 10,93% | 1,55 | 1,28 | 0,67 | 1,57 | 0,59 |
| 80 | 214 | 438 | 631 | 1069 | 51 | 12,22% | 1,65 | 1,48 | 0,69 | 1,83 | 0,62 |
| 90 | 239 | 437 | 746 | 1183 | 44 | 13,66% | 1,75 | 1,72 | 0,72 | 2,12 | 0,65 |
| 100 | 267 | 437 | 874 | 1312 | 36 | 15,27% | 1,86 | 1,99 | 0,75 | 2,46 | 0,69 |

The baseline scenario of Table 1 suggests that with the initial value for *K* and the chosen parameters, the capitalist road to communism is economically feasible. But actually reaching communism will take more than 100 years, since one century later labor demand is still almost half of its initial value. Even so, there are major doubts about the social and political viability of the transition. We address these in section 5. In the remainder of this subsection, we discuss various aspects of the simulation.

Somewhat artificially, the *CAP* and *MS* versions of the model shown in 3.1 are assumed to have the same constant savings rate and have the same technological and labor supply parameters. Therefore, the two behave much in the same way, as Table 1 shows. Only the tax rates of *CAP* and *MS* for financing *G* differ. As equations (5) and (5') show, *CAP* finances $nG$ from a uniform tax on total income, while *MS* finances part of $nG$ from the total return on state-owned capital $rK$ and uniformly taxes $wL$, therefore $t_w$ lies below $t$ in the Table. The similarity in behaviour occurs because labor demand in the model is independent of the institutional form, either *CAP* or *MS*. As a result, the equilibrium level of basic income in each system follows from the equality of labor supply *L* in *CAP* and *MS*, and therefore from the equality of the disincentive terms, respectively $\delta G/wQ(1-t)$ and $\delta G/wQ(1-t_w)$ in



equations (6) and (6'). It should be noted that the development of $G$ and $t$ or $t_w$ is quite sensitive to the behavioural parameter $\delta$ which we assumed to be constant over time.

In Table 1 and in other simulations shown in section 5 below we chose to have $G$ initially at one half of per capita income in *CAP*. To achieve this, we fixed the sensitivity parameter of labor supply at $\delta = 25$. The choice is empirically motivated. At the start of a possible transition process, we want the income share of social expenditure and basic income $nG/Y$ (and thus the tax rate, see equation (5)) to be around 50% in order to roughly represent the ruling circumstances of a moderately affluent capitalist welfare state. Having assumed that $nG$ not only comprises the unconditional cash grant of basic income for the population of such a welfare state, but also - and in considerable part - state expenditures on education and health care, as well as certain funds ensuring equal access to vital provisions of transport and housing (basic income in kind, so to speak), we realistically need to have $nG$ initially taking up around one half of national income $Y$. Since the *CAP* version of the model is the one for which we wish to ensure some empirical plausibility, the initial (and somewhat higher) value of $G$ for the *MS* version necessarily follows.

As mentioned in 2.3, the composite interpretation of $G$ necessitates specifying the shares of unconditional basic income in cash (*UBI*) and of the other social expenditures (*SE*) in a longer timeframe. To simplify, we assume that these shares remain constant over time and that the share of *UBI* in $G$ is one half. Then in $T = 0$, the unconditional part of $nG$ is 25. It amounts to 25% of per capita income. This may be close to weak abundance in an affluent Nordic welfare state. We thus assume $\frac{1}{2}G = UBI = SE$ in all periods. This means that during the transition to communism, both *UBI* and *SE* will be growing faster than per capita income. In terms of the normative criterion of a 'shift from contributions to needs', our assumption about the



proportion of *G* going to *UBI* fits in well with Marx's own remarks about the growing importance of funds for education and health in the *Critique of the Gotha Programme*.[30] It also reflects that the expenses of the welfare state (*SE*) and unconditional cash transfers (*UBI*) are in competition for resources, which as we saw earlier poses a problem when the economy is still below weak abundance. And at least for sectors such as healthcare and education, our constant-share assumption seems roughly tenable.

To conclude this discussion, a comment on the measures of output and capital goods used in the SA model is in order. Since total income *Y* is treated implicitly as a single good at a price of unity these are nominal measures, so the model does not reflect changes in real income over time. It hence fails to give adequate quantitative indications of the state of material abundance. For example, comparing income at two periods, say $Y = 100$ in $T = 0$ and $Y = 193$ in $T = 70$ in Table 1, a fundamental reality of technology - saving costs and introducing new products through innovation - remains hidden from sight. Depending on the speed of changes in productivity, later periods may represent a whole new array of inputs and outputs which are bound to be sold at much lower prices. Especially when looking at the possible uses of AI in widely different fields such as medicine, media entertainment or production of new materials, this point should be kept in mind, as we shall also see next.

### 4.2 Comparing different rates of saving and technical change

For purposes of social and policy evaluation it is important to compare how the SA model behaves under different rates of saving and investment *s*, and under different productivity parameters *A* and *B* governing labor and automation capital respectively. These two parameters determine the nature of technical change - in particular the way in which automation proceeds - whereas the savings rate determines how rapidly a pattern of technical change described by given values of $A$ and $B$ is being implemented over time. We first compare two economies which are the same in all respects except saving, and then look at two economies with different patterns of technical change, all other things equal. Our discussion starts with a striking fact. In both of these comparisons, nominal income $Y$ has the same value in the two economies, as our simulations in the Tables 1 of sections 3.1 to 5.2 and 5.3 show. However, what remains hidden is that in these Tables, the two economies under comparison will have different levels of real income and different compositions of goods and services, as explained in 4.1 above.

---

[30] Marx, 1975, 567. Marx's 'funds for those unable to work' however, are here replaced by unconditional basic income.

Consider two economies E1 and E2. E1 has a higher rate of saving than E2. In E1, the given technology has been implemented on a larger scale than in E2, and therefore has gone further in replacing labor by automation capital. In E1, prices of inputs and final products are lower on average, implying that it has more real income than E2. The invariance of nominal $Y$ in this first case of comparison is now explained by looking at a central property of the SA model. It says that the allocation of capital goods of the two types and of labor is regulated by the equality $w/r = A/B$. Given that the technology is the same in both economies, E1 and E2 have the same rates of return on labor and capital. Comparing values at any moment after $T = 0$, this implies that in E1 the positive difference in cost of capital goods at the ruling rate of return to capital must equal the negative difference in cost of labor at the ruling wage rate. Thus $rdK = -AdL$, where *'d'* is the numerical difference of $K$ or $L$ between E1 and E2. Therefore, $AL + BK$ is the same in both economies. Now, from (9) and the reduced forms of $w$ and $r$ (see 3.1) it follows that $Y = h(AL + BK)$, where $h = (\alpha^{\alpha}(1 - \alpha)^{1-\alpha}) / B^{\alpha}$ is a constant: $\alpha$ and $B$ are the same in E1 and E2. This confirms that $Y$ is invariant under the difference in $s$ between the two economies. Given this invariance, the value of fixed capital



$K_2$ is the same as well, since $K_1 = \alpha Y / r$, from (2). Thus, the additional capital used in E1 consists entirely of automation capital. Moreover, E1 has a higher capital share $rK/Y$ than E2. Moving now to income distribution, we see that E1 has a higher level of basic income $G$ in order to match its labor supply to the lesser amount of labor demanded, in comparison with E2. This follows from the labor supply function in (6), since the wage rate $w$ and the quality of labor $Q$ are the same in the two economies. Given the invariance of $Y$, again, the tax rate $t = nG/Y$ is higher in E1 than it is in E2.

The implications of this first case of comparison for the transition to communism are clear. If other things remain equal, more investment appears to be beneficial. It speeds up the transition on the labor criterion *M2*, as well as on criterion *M1* in terms of real income, though not in terms of nominal income, due to the invariance property of $Y$. Finally, the higher rate of saving in E1 brings more progress along criterion *M4*. It guarantees a higher $nG/Y$ than would be viable in E2, and therefore better serves the transition to communism by achieving a more marked 'shift from contributions to needs'.[31] This straightforward conclusion should be qualified on two counts. First, more investment in automation also increases the capital share, as noted above. On our criterion *P5*, this is how the capitalist road to communism will be likely to proceed. But it presents a larger risk that the transition mechanism of raising basic income to let the labor supply match a falling demand for labor can be undone by the stronger influence of economic inequality on politics. Secondly, more investment means more output of consumption goods and throughput of capital goods. The higher abundance in terms of available goods and time off from paid work existing under E1 (relative to E2, the economy with less saving) may therefore meet hard ecological constraints. This consequence was far less apparent in Marx's time.

---

[31] The consequences of attaining more economic growth due to a higher rate of saving in the SA model are different from the trade-off between the 'growth-oriented' (M1) and 'free time-oriented' (M2) choices of a transition path identified in *Capitalist Road* on the basis of a more primitive economic model (see the section on the Laffer Curve in Figs.1 and 2, 647-652).

We now turn to the second synchronic comparison of two economies with different productivity parameters $A$ and $B$, now called E3 and E4, which have the same rate of saving. E3 is the most productive economy of the two. Suppose for simplicity that its automation capital is more efficient - it has a higher value of parameter $B$ at the given time of comparison than E4 has, and has the same value of $A$. To assess the difference between E3 and E4 we note that their capital stock $K$ is assumed to have the same value, even though the stocks of E3 and E4 have different compositions, due to the consequences of the higher value of $B$ in E3. At the same rate of saving, (10) and (7) then imply that nominal income $Y$ also has the same value in the two economies. Again however, and to the extent that the higher productivity of automation capital generates lower prices, there will be more real income in E3 than in E4.

To continue our assessment, the expression for $Y$ used above is helpful. Once more we have $Y = h(AL + BK)$. But now $h = (\alpha^\alpha (1 - \alpha)^{1-\alpha}) / B^\alpha$ is no longer a constant, because $B$ is higher in E3 than in E4. E3 has a higher rate of return $r$ on capital, since $r = hB$, and E3 has a higher value of $B$ at the same value of $\alpha$. Given the invariances of $Y$ and $K$, two things follow. First, E3 has a larger capital share. Second, $K_1$ will be smaller in E3 than in E4, from (2). In turn, $K_2$ must be larger to make up for the difference, and $L$ will be lower as a result. So in E3, more labor has been replaced by automation capital than in E4. And, following the same reasoning as above, E3 also has a higher $G$ and $t$ than E4. The implications of this second case of comparison for the transition to communism are very similar. If other things remain equal, having a more productive stock of automation capital appears to be beneficial in terms of the



transition criteria *M1*, *M2* and *M4*. The two caveats mentioned above concerning the larger

dangers of inequality and (to a lesser extent) the possible limits on growth also apply.

This comparison between two differentially productive economies with the same rate of

saving is not the only relevant one. There are other cases in which the more productive of the

two in terms of automation - a higher value of $B$ - could score higher in terms of progress

along the labor dimension *M2*, but have less progress on the distributional criterion *M4*, in

terms of $G$ and $t = nG/Y$. As will be seen in section 5, this is a distinct possibility if the other

economy has a higher productivity of labor, as shown by parameter $A$. To the extent that

choices of technology as these are on the social and policy agendas of countries undergoing a

transition process, trading off possible developments along *M2* against those along *M4* will be

a normatively relevant issue. And as we will also try to show in the lengthy section now

following, such choices of technology open up a host of other issues which serve to clarify the

promises and problems which a conception of a capitalist road to communism brings up.

## 5. The transition to communism: a speculative scenario

### 5.1 Technical change and institutional reform

The original article of *Capitalist Road* and the replies to critics in the two symposia (van der

Veen and Van Parijs, 1986b, 2006) did not consider technological determinants affecting the

timescale of the capitalist transition to communism under the pressures of automation, nor did

they consider reforming welfare capitalism in the direction of market socialism to mitigate the

pressures of inequality. Both pressures have now become stronger, hence we address the

interplay between technical and institutional change in this final section. We believe that a

piecemeal but comprehensive reform of welfare capitalism towards some form of liberal-

egalitarian market socialism offers the most realistic chances for embedding unconditional



basic income - the key policy of the transition to communism - into an economy with strong labor-saving tendencies. We do not go into detail about the exact features of such a major reform process and how these features affect or are affected by growth and technology. We rely instead on listing some of the main ideas in Thomas Piketty's useful conception of participatory socialism in Chapter 17 of *Capital and Ideology*, supplemented by recent work of others. Those ideas, while limiting capitalist property rights in various ways, leave room for the market and can be seen as elements of a liberal-egalitarian market socialism which satisfies normative and institutional prerequisites for using the instrument of unconditional basic income to travel the capitalist road towards communism. A large part of Piketty's conception can be represented by a *continuum* running from welfare capitalism - as that system presently exists in its state of being under threat by inequality of chances and material conditions - to this form of market socialism.[32] The continuum has five dimensions.

---

[32] As mentioned in the introduction, we do not go into issues of environmental and international justice for reasons of keeping a focus on a national society. We thus leave out the global wealth and carbon taxes mentioned by Piketty. Of course this is not to deny that such measures would increase the likelihood of a transition to communism in national societies.



Textbox 2. A reform continuum: welfare capitalism to market socialism[33]

> (1) stronger redistribution or 'circulation' of personal wealth, including one-time endowments on reaching adulthood, supported by steeply progressive taxation of capital and inheritances.
> (2) more equality of sharing the fruits of human capital (by compression of wages and non-pecuniary benefits) supported by progressive taxation of income and consumption, regulation of work conditions, and limitation of high-end remunerations.
> (3) socialization in the form of public-private tax-financed 'funding missions' for renewing housing, transport and infrastructure, spearheading research and development, managing the energy transition, adapting to climate change and redirecting technology.
> (4) more equality for welfare-state provisions in education at all levels, health, childcare and housing, transforming these provisions into 'basic services' in a social investment strategy aimed at preventing lifetime risks.
> (5) more co-determination and participation of workers, consumers, and environmental interests in firms and public organisations, supplemented by measures to reduce political inequality in representative bodies, such as democratic vouchers and regulation of campaign financing.

As Piketty and others have argued, these and other moves in the direction of socialism are guided by historical development and by an ideology more or less following the principle that just inequalities must be to the benefit of the least advantaged.[34] These moves are likely to be contested on fine points of normative justification as well as local application, and can only be expected to be taken erratically. Therefore, it may be difficult to judge how welfare capitalism might be moving toward market socialism. In our scenario, we leave this largely open. We only suppose that institutional reforms of the kinds just listed are in the process of formation and take place over time to help along a transition to communism, rather than being posited as theoretical preconditions for a viable transition to start in the first place. Moreover, the extent to which these reforms are carried through in periods of rapid automation is likely to influence the pattern of innovation underlying particular instances of the transition, In this respect, the different dimensions of the continuum hang together to some extent.

---

[33] Dimension (3): for the importance of funding missions for increasing the effectiveness of government intervention in capitalism see Mazzucato, 2018, on the need for redirecting technology see Korinek and Stiglitz, 2020 and prominently, Acemoglu and Johnson, 2023, Ch. 11. Dimension (4): on social investment see Martinelli and Vanderborght, 2022.

[34] Piketty (2019, 968-9 and notes 1 and 2) provides a guarded indication of adhering to a Rawlsian framework of social justice, even though he does not underwrite our favored version of the difference principle defined over income-leisure options (see note 2 above). In Piketty's own income proposals, which are covered broadly in dimension 2 of Textbox 2 above, a kind of basic income taking the form of an earned income tax credit also figures, the average amount of which is tied to around 16% of national income per adult. Piketty thus rejects the alternative of a fully unconditional basic income, wanting to link 'basic income to a vision of the just society based on the wage relation and the right to work and unionize.' (Piketty, 2019, 1002-3 and note 84). Despite this rejection of a highly specific detail in Piketty's grand overview of elements of participatory socialism, we can profitably remain using the overview as a broad guideline for identifying the kind of social reforms that may be needed to pave the capitalist road to communism.

Our scenario thus assumes that technological and institutional changes be roughly aligned. Looking at the present state of globalisation and geopolitics in the world, technical change develops autonomously for some part. Institutional alignment then consists of adapting in the right ways to let the transition process to stay on course, or to regain stability in the face of periodically occurring disruptions. For another part, technology is wide open to social choices. Awareness of this motivates reshaping the goals and concrete purposes of innovation to better suit the interests of the workers and other stakeholders. As Daron Acemoglu and Simon Johnson forcefully argue (2023, Ch 9 and 11) it calls for redirecting technology away from profit-maximizing powerholders to making technical change led by automation more 'labor-enabling' and less - or not at all - 'labor-replacing'.[35] We incorporate some of their insights below. But we strongly agree with Anton Korinek (2023, 32) that even if this kind of redirecting succeeds, a *post-labor world* generated by a technology driven by AI in combination with other general-purpose innovations still remains a realistic possibility. On the way to a future of redundant labor, automation might or might not go together with steering technical change in order to enable as much as possible the human work that remains engaged in production (Korinek and Stiglitz, 2020, Working Paper for the NBER Economics of AI Conference, accessed from Korinek's webpage). By looking at these possibilities, the transition to communism can be further clarified. We will do so with some simulation illustrations.

---

[35] Note however that Acemoglu and Johnson do not appear to be in favor of Marx's vision of a realm of freedom without labor, a vision which they do not discuss. They seem to want a more or less indefinite future in which well-paid workers spend a substantial part of their time doing meaningful jobs, and they reject unconditional basic income as a way of softening the impact of automation (pp. 416-7). See also Susskind, 2023. In particular Acemoglu, in a podcast discussion with Korinek, says it is just as well that a jobless future is highly unlikely to emerge, because he thinks such a future would be shaped by the decisions of a small elite of technology owners that controls automation and creates a society of useless second-class citizens. Handing out a basic income would only serve to perpetuate the inferior status of these citizens, see Centre for the Governance of AI (GovAI), (2023), Daron Acemoglu Webinar on Power and Progress. Our 1000-Year Struggle Over Technology and Prosperity, October 10, 42'-43' at https://www.youtube.com/watch?v=jz37eST8hdg

*5.2 A scenario in two phases*

The following two-phase scenario lays out what one might conservatively expect to happen within the SA model if automation continues to unfold.[36] Automation initially develops at a relatively slow pace and is quickened as the transition process takes hold. Institutional reforms allow income to be distributed more in favor of workers without capital income over time, even though capital owners benefit to a somewhat greater extent. There is a preliminary *first phase* which lasts 10 years (up to $T = 10$). The economy is not yet net labor-saving and barely at weak abundance, but there is a substantial partial basic income. One main concern is securing the co-existence of egalitarian welfare state provisions, especially in education, with the growing requirements of unconditional basic income.[37]

The scenario assumes that the interplay of unevenly proceeding technical change and institutional reform will not produce a smooth growth path. Due to surging innovations in automation together with breakthroughs in other general-purpose technologies (for pertinent descriptions see Suleyman 2023, Chapter 4) the transition process enters an *era of disruption* which breaks the relatively calm steady state of the economy's first phase. This period lasts nine years after $T = 10$. Its details are left outside the purview of the scenario. Such a long period will be quite turbulent. However, in the optimistic exercise in which we are here engaging it should end auspiciously. So the economy again settles down to a steady state in a *second phase* of the capitalist transition to communism, in which labor is replaced by automation. This can happen in different ways and to different extents. Some progress along the five dimensions of the reform continuum of Textbox 2 will be needed to achieve the second phase, but the degree to which such progress occurs within the era of disruption is uncertain. In our scenario we therefore focus on two possible versions of the second phase. They show the trade-off between the Marxian transition dimensions *M2* and *M4* noted above in 4.2. In the rest of this section, we identify the version which arguably represents the most appealing form of the transition to communism. Our treatment of the two phases of the whole scenario rests on two conjectures.

---

[36] Our model assumptions are more conservative than the ones that Korinek mentions: artificial general intelligence pervading all economic domains and implementable by massive savings on costs resulting in $L = 0$ within either twenty years or (in an 'aggressive scenario') five years (Korinek, 2023, p. 32).

[37] On the need to diminish educational inequality between groups leaving school at early or late ages see Piketty (2019, 1007-1011), and on the need to finance and shape specific programs of vocational training to increase the digital skills of workers across education levels, see Acemoglu and Johnson (2023, 331-3 and 417-8).

*Conjecture 1*: when the rate at which automation becomes more productive - as measured by parameter *B* - increases, the rate at which labor becomes more productive - as measured by parameter *A* - also increases to a greater or lesser extent. More efficient AI applications can in principle boost the accuracy, speed and creativity by which tasks are performed by humans before they finally get replaced. Thus, when automation speeds up, the growth rate of labor $g_A$ tends to rise together with the growth rate of automation capital $g_B$. However, depending on choices of technology, the productivity of automation capital may either grow much faster than that of labor, or the two factors become more productive at more equal rates of change.

*Conjecture 2:* when technical change of the labor-replacing kind (in either of the two broad variants just mentioned) is being introduced as innovation proceeds more rapidly, automation costs will fall, applications of automation in small and middle enterprises will increase and expectations of profitability will proliferate. As a result, the rate of investment goes up.



The first phase of our scenario is shown in Table 2. It has a transition path with a moderate rate of saving $s = 0,15$. The annual productivity of labor and automation capital grow equally at a modest rate of 0,7 percent. The second phase is presented in 5.3 below. It will show possible variations in the ways that $g_A$ is linked to $g_B$. Our numerical examples proceed from the initial values of the baseline example in Table 1.

Table 2 First phase of the scenario: no net labor-saving automation

$s = 0,15$, $g_A = g_B = 0,007$; $T = (1 - 10)$

| T | Y | $K_1$ | $K_2$ | K | L | r | w | $G_{CAP}$ | t |
|---|---|---|---|---|---|---|---|---|---|
| 0 | 100 | 500 | 0 | 500 | 75 | 5,00% | 1,00 | 0,50 | 0,50 |
| 5 | 104 | 507 | 18 | 525 | 75 | 5,13% | 1,03 | 0,52 | 0,50 |
| 10 | 108 | 515 | 36 | 551 | 75 | 5,27% | 1,05 | 0,54 | 0,50 |

In this preliminary phase, the institutions of basic income are in place. Technical change driven by automation does not yet lead to widespread displacement of labor. The transition is not properly underway, for the labor and distributional criteria *M2* and *M4* are not satisfied, as shown by columns L and t, (with $t = nG/Y$) in Table 2. In these first ten years, the level of unconditional basic income (assumed to be initially at one half of $G$, see 3.1) is relatively low and barely meets the subsistence requirements of weak abundance. Thus, the bargaining power of workers conferred by the exit option of basic income is still not that strong, and problems of ensuring good quality of labor within firms and public organisations may exist, even if automation improves the productivity of workers (see Furendal and Jabari, 2023). Martinelli and Vanderborght (2022) discuss specific opportunities of reforming the welfare state which are important in the first phase of our scenario, in the course of arguing for institutional complementarity between programs of 'social investment' and basic income.[38]

---

[38] Martinelli and Vanderborght (2022, 48) convincingly argue that a basic income policy need not be incompatible with the social investment strategies advocated in European welfare states, as is often maintained: "…there should be no presumption that labour market disengagement is the goal or indeed the likely impact of the introduction of all forms of BI. Core elements of the "activation" agenda – such as (voluntary) skills



Even though the gross share of labor income falls slightly, workers fare reasonably well on their net wage and basic income.[39] In our scenario however, the nine-year era of rapid technical change disrupts this peaceful phase after $T = 10$. For example, sudden bursts of innovation are not well-aligned with existing institutional arrangements because of lack of information, financial incentives of shareholders, problems of (self-)regulating decisions in powerful firms in which AI-innovations are being introduced quickly, and unevenly distributed opportunities for sectors to adopt promising innovations. How these events play out in adjustment processes depends on the force of reforms to provide pressures in the direction of market socialism of the kinds listed in Textbox 2. Such pressures may be more or less strong and will affect innovation in different ways.

*5.3 Economic and normative concerns of the transition: second phase*

As mentioned in 2.3, the key policy of setting the level of basic income to prevent involuntary underemployment is also required by the principle of social justice that requires maximizing the income-leisure options of the least advantaged position in society. That position is tracked by the level of unconditional cash dispensation, a level which will have to rise significantly when technical change speeds up, as the numbers in Table 3 below show. We optimistically suppose that the normative motivations behind the transition to communism will survive the era of disruption to a greater or lesser extent. As indicated above, we assume that this era can give rise to two alternative versions of the second phase of our scenario, each of which starts in $T = 20$. By that time, the transition is underway. All six of its criteria *M1-M4*, *P5* and *P6* are

---

development programmes and public employment services are entirely consistent with partial BI schemes." We also think that duly crafted social investment programs in an 'extensive welfare state' will complement full basic income, and can therefore be regarded as components of the non-UBI part of *G*. Such reforms fit in dimension 4 of Textbox 2.

[39] From Table 2, income as measured by an initial value of *w(1 - t)L/ n + G* stands at 0,875 in $T = 0$, compared to a value of 0,887 in $T = 10$. The labor share (1 - *rK/Y*) - stands at 0,75 in $T = 0$ and at 0,739 in $T = 10$.



fulfilled.[40] The alternative versions of the second phase are displayed in Table 3 below. Both are in line with Conjecture 1. The numerical examples for $T = 20$ and onwards are obtained by extrapolating the row of initial values at $T = 0$ of Table 2, using the selected values of saving and technological parameters which are given below. Each version of the second phase thus presupposes a specific history of the era of disruption, which is supposed to end with the variables shown in the first row of Table 3a, or alternatively with the ones shown in Table 3b's first row. Depending on what has happened during the era of disruption, then, our scenario unfolds in two different ways, to be discussed in what follows next.

In what we call the *slow road*, displayed in Table 3a, the rate of saving is $s = 0{,}25$, reflecting higher incentives to invest in processes of more rapid technical change than those obtaining in the first phase of the transition in which $s = 0{,}15$, in agreement with Conjecture 2 (see 5.2). In this slow version of the second phase, parameters $A$ and $B$ increase by 1,8 percent annually. It reaches a notional post-labor state in which $L$ drops to almost zero at $T = 68$, $G = 1{,}66$ and $t = nG/Y = 0{,}88$. In the alternative *fast road* of the second phase, the rate of saving likewise is at 0,25. Here, parameter $B$ of automation capital strongly rises by 2,5 percent, whereas parameter $A$ of labor productivity grows by only 0,7 percent. As a result of these differences, the stock of automation capital $K_2$ is much higher in the fast scenario. Therefore $L$ falls swiftly to almost zero in $T = 51$, with $G = 1{,}22$ and $t = nG/Y = 0{,}77$. At that same period, as Table 3a shows, the slow version has $L = 19$ , $G = 1{,}33$ and $t = nG/Y = 0{,}84$. In this version, the economy is still dependent on labor to a significant extent but also has a significantly higher level and share of basic income. The two versions of the second phase thus exemplify the trade-off between the labor criterion *M2* and the distributional criterion *M4* studied in 4.2.

---

[40] Regarding *M1*, real income increases more than the figures for *Y* indicate; regarding the Piketty criteria *P5* and *P6*, calculating values for *rK/Y*, *r(1 - t)* and *g* show that in both versions, the capital share increases and the net rate of return on capital exceeds the growth rate of income.



Due to the invariance property of $Y$ discussed in 4.2, they perform equally on the nominal income criterion *M1*. But of course the composition of goods in both economies is different, as is further discussed below.

Table 3. Net labor-saving automation in the second phase of the transition scenario: a slow road (3a) and a fast road (3b)

3a. The slow road: $s = 0{,}25$, $g_A = g_B = 0{,}018$, $T = (20$ to $68)$

| $T$ | $Y$ | $K_1$ | $K_2$ | $K$ | $L$ | $r$ | $w$ | $G_{CAP}$ | $t$ | $w(1-t)$ |
|---|---|---|---|---|---|---|---|---|---|---|
| 20 | 118 | 452 | 333 | 786 | 51 | 6,53% | 1,31 | 0,83 | 0,70 | 0,386 |
| 30 | 130 | 434 | 491 | 924 | 41 | 7,47% | 1,49 | 0,99 | 0,76 | 0,357 |
| 40 | 142 | 417 | 648 | 1065 | 30 | 8,54% | 1,71 | 1,14 | 0,80 | 0,336 |
| 50 | 157 | 402 | 809 | 1211 | 20 | 9,76% | 1,95 | 1,31 | 0,84 | 0,319 |
| 51 | 159 | 401 | 825 | 1226 | 19 | 9,89% | 1,98 | 1,33 | 0,84 | 0,318 |
| 60 | 174 | 389 | 977 | 1366 | 10 | 11,16% | 2,23 | 1,50 | 0,86 | 0,307 |
| 68 | 189 | 380 | 1119 | 1498 | 1 | 12,42% | 2,48 | 1,66 | 0,88 | 0,299 |

3b. The fast road: $s = 0{,}25$, $g_A = 0{,}007$ $g_B = 0{,}025$, $T = (20$ to $51)$

| $T$ | $Y$ | $K_1$ | $K_2$ | $K$ | $L$ | $r$ | $w$ | $G_{CAP}$ | $t$ | $w(1-t)$ |
|---|---|---|---|---|---|---|---|---|---|---|
| 20 | 118 | 408 | 378 | 786 | 60 | 7,24% | 1,02 | 0,71 | 0,60 | 0,406 |
| 30 | 130 | 372 | 553 | 924 | 48 | 8,71% | 1,02 | 0,85 | 0,66 | 0,351 |
| 40 | 142 | 340 | 725 | 1065 | 30 | 10,49% | 1,03 | 1,02 | 0,71 | 0,296 |
| 50 | 157 | 311 | 900 | 1211 | 4 | 12,62% | 1,04 | 1,20 | 0,77 | 0,244 |
| 51 | 159 | 308 | 918 | 1226 | 1 | 12,86% | 1,04 | 1,22 | 0,77 | 0,239 |

We now present a two-part argument as to why the slow road has greater normative appeal than the fast one. In the fast version, a future without any paid work materializes far more rapidly than in the slow one. But this could be bad for those who undergo it. The first part of our argument to this effect notes that the slow road is superior on the distributional criterion *M4* in all periods of comparison up to $T = 51$. It has a larger $G$, hence - with $Y$ at the same value in both versions - a larger $t = nG/Y$.[41] The slow road also arguably brings higher labor quality to workers, so it is also superior on the labor quality criterion *M3*. This is not only because a higher basic income gives workers a stronger exit position, but also and far more importantly, because the technology of the slow road increases labor productivity (measured by parameter *A*) at a much faster rate and thereby gives labor a more important role in the production process with more varied and responsible tasks that replace more boring and less intricate ones.[42] So, on the slow road there will also be more at stake for workers involved in co-determining the organisation of tasks. The bargaining position conferred on them by a higher basic income will not only be better than in the fast track. It will also be deployed to address a whole range of comparatively more important issues of work design, due to the specific pattern of technical change exemplified by the slow road.[43]

---

[41] At first sight, the slow road is better for the workers than the fast one if one only looks at the change in the labor share of income *(1 - rK/Y)*. Between $T = 20$ and $T = 51$, the labor share goes from 0,599 to 0,237 in the slow version, and from 0,518 to 0,008 in the fast one, as calculated from Tables 3a and 3b. But this is misleading. A more relevant measure is the basic income augmented with net income of people with average work and no capital income, see note 38 above. Here also, the slow road does better than the fast one. But the difference is quite small and anyhow is tracked well by the level of basic income. For due to the rise of $t$ and the fall of $L$ in both versions, the personal income component $G$ is by far the largest.

Of course, against these desirable features one must weigh in the fact that the slow road keeps people occupied in paid work for no less than eighteen years longer than the fast road does. However, as Tables 3a and 3b show, at the start of the second phase, less labor is used on the slow road than it is on the fast one, for at $T = 20$, $L = 51$ against $L = 60$. Because the decrease of $L$ on the fast track is accelerating, there is a turning point at $T = 40$, at which time both versions have the same labor volume of $L = 30$. To illustrate in terms of a hypothetical working week, suppose $L = 75$ at $T = 0$ equals five weekdays of seven hours on average. Then $L = 30$ at $T = 40$ corresponds to two days of work totalling fourteen hours. This is close to Keynes's utopian image of the 'leisure economy' with its fifteen hours of weekly work in

---

[42] A warning is in order here. When automation improves the productivity of labor and parameter *A* increases, this does not logically entail improvement in the quality of labor. Furendal and Jabari, 2023, section 5.1, give contemporary examples of how lower-skilled workers can be made more productive by automated devices that cut some of their tasks, leading to a more repetitive work menu, while surveillance devices and tracking algorithms enable employers to compel these workers to complete the menu at increased speed. We here assume that such quality-decreasing changes will not be adopted in either of the two versions of the second phase because of workers' strong exit position under a high level of basic income.

[43] The higher average quality of work *Q* of the slow road should also generate a higher basic income than is recorded in Table 3b, due to its positive effect on labor supply, see equation (6) in Textbox 1. This because in our simulations *Q* is a parameter which grows at the same rate irrespective of the kind of technical change.



*Economic Possibilities for our Grandchildren* (Keynes, 1933/1963). After the turning point at $T = 40$, the fast road quickly stops using labor, going from $L = 30$ to $L = 1$ in just eleven years, at which time $L = 19$ on the slow road. Since in both versions $Y$ is the same, this represents a 19-fold advantage of the fast track in terms of the customary measure of productivity per hour at $T = 51$. Does that provide sufficient grounds for preferring the fast track of the transition's second phase despite the advantages of the slow road just mentioned?

The answer is no, according to the second part of our argument. Up to now we compared values of the transition criteria *M2 (L)* and *M4 (nG/Y)* of the two versions at given time slices. We also made informed guesses about the quality differences in work processes. Now we follow up by reflecting on what the history of events at any particular time slice of the second phase may have been. This brings out likely background facts which not only confirm that the slow version offers better labor quality within firms and public organisations - which directly matters as long as labor is being done for money - but also offers a better quality of many of the varied activities in the 'realm of freedom' during $T = 20$ and $T = 51$, the period in which the two versions are being compared in Table 3. Such activities run from recreation and travel, leading a family life in the household to formal and informal political participation including the execution of communal projects. On our general understanding of what raises the quality of purposeful activities set out in 2.3, most of the activities just listed will most probably have a better quality than they would have on on the fast road, being supported by a higher basic income. In addition, and more importantly, such free-time activities will benefit to different degrees from the knowledge and skills acquired in performing more demanding semi-automated tasks in the sphere of paid labor, especially when those work tasks are new, or can otherwise be improved by economizing on time-consuming and boring sub-tasks.



Reasoning backwards to the era of disruption, this richer offering of education - both in theoretical and practical fields - must already have been present prior to the emergence of the slow road (see Textbox 2, dimension 4). Such educational provisions will have to be in place if policymakers are to redirect technology effectively (Textbox 2, dimension 3) at the start of the second phase, and make it possible for firms and public organisations to undertake the technical changes that boost the productivity of automation capital and better schooled labor in tandem, rather than making labor redundant as quickly as possible. Automation of this labor-augmenting type is not only useful for individuals inside the sphere of labor, as Acemoglu and Johnson strongly emphasize from a historical perspective (2023, Ch. 9). It also makes many instrumental non-labor activities easier to perform, in the sense that it leads more quickly to the desired results while leaving time for intrinsically more rewarding sub-goals, especially when both workers and consumers are assisted by personal AI's waiting to be prompted. In the slow version, members of a better educated workforce - and of a more sophisticated 'consumption force', increasingly - will also be more likely to autonomously choose results of spending their time for goals they consider desirable in the first place. This general effect is important in view of automation's enhanced power to influence preferences and decisions. Next, within the social conditions of the slow version sketched so far, individuals are more likely to be aware of this pervasive influence than they would be in conditions corresponding to the fast road of the second phase. That awareness will be enhanced to the extent that the slow road has more developed procedures of co-determination in different fields outside of parliamentary democracy (see Textbox 2, dimension 5). Making possible and protecting more autonomous responses to automation by workers and consumers will in turn invite more effectively safeguarded regulation of media and commerce, and create more opportunities to articulate and take democratic decisions than will presumably be the case on the fast track. Finally, the more equal distribution of power between 'Capital and the



People' which almost certainly is needed to achieve all these technological and behavioural effects can also be thought to underlie the slow road of the transition's second phase.

*5.4 The two versions of the scenario on the way to the post-labor stage.*

On the liberal-egalitarian interpretation of communism we hold, then, the slow road of the scenario is the one to be recommended so far, after comparing the numbers in Tables 3a and 3b of two versions between $T = 20$ and $T = 51$, and reflecting on the social and political conditions needed to make one or the other version more likely to be a genuine transition to communism. So far, the slow road appears to score higher on different desiderata indicated by the Marxian concept of alienation which we discussed in 2.2 and 2.4. For in our admittedly speculative argument, the slow road promises to enable a progressive suspension of the distinction between work and free time at a higher level of individual autonomy, social connectedness and well-being. To really have these fine qualitative characteristics, we argued that the slow road will also have had a more market socialist history than the fast one, on several of the five dimensions described in Textbox 2 of 5.1. And this in turn makes it more likely to think that the normative motivation to keep on raising the level of $G$ so as to adjust $L$ downwards at the required rate shown in Table 3a will also be more alive than it is in the case of the fast road in Table 3b.

This last point can also help assess the two versions between $T = 52$ and $T = 69$, the later period of time in which the slow road is still moving toward the post-labor stage and the fast one is already seventeen years inside it. For those years, we have numbers for the slow road, but not for the fast track, because events in this version of the transition are somewhat of a mystery after $L$ has reached zero. To explain why, look at the equations in the *CAP*-format of Textbox 1 (in 3.1). In the post-labor stage, we must set *A, L* and *w* at zero. The SA model



shows that $t$ is then free to vary. Thus in the post-labor stage, the policymakers of the fast track are at liberty to set the uniform tax rate on income within wide limits. The lower limit is the one generating a level of $G$ ensuring a weakly abundant economy. As defined in 2.3, and given a number in 4.1, one half of per capita income is the level of $G$ ensuring weak abundance (i.e. $Y/n$ in $T = 0$, at $t = 0,5$, see Tables 1 and 2, first row). The upper bound of the tax rate must be $t = 1$, the level of tax embodying strong abundance, in which all income is distributed equally: $tY = rK = nG$. This last level is also the *only one* prescribed by the *MS*-format of the SA model for the post-labor stage. For in that format, the part of $nG$ financed by a tax on wage income $t_w wL$ has disappeared and we have $nG = rK = Y$ (see Textbox 1). Given the relatively capitalist history of the fast version - a history leading to widespread investment in uses of automation that will eliminate labor sooner rather than later - it is reasonable to think that the *CAP*-format rather than the *MS*-format will be the one in force, and that once the tax rate is free to vary, the proportion $t$ of per capita income financing basic income and egalitarian welfare provisions will be far closer to the weak abundance ratio of one half than to the strong abundance ratio of unity.

On this hypothesis, the fast version of the capitalist road does not really issue in a communist post-labor society. But of course, one can still continue comparing the two versions. If our thinking is right, then once again the distributional criterion $nG/Y$ and the labor criterion $L$ must be traded off between the fast and the slow versions in the later period of comparison. For example, consider the trade-off at $T = 52$, just when the fast version has entered the post-labor stage, not shown in Table 3b. Both versions then have $Y = 160$ and $K = 1241$. The slow road has the high levels of $G = 1,35$ and $t = 0,84$ at $L = 18$, whereas post-labor, $G$ and $t$ on the fast road will have levels approximating weak abundance, at close to $G = 0,8$ and $t = 0,5$.[44]

---

[44] In $T = 52$ private owners in the fast version will reap a net reward of about 6,4 per cent, calculated as $r(1 - t) = 0,0645$ with $r = Y/K = 160/1241$ and $t$ (if set at its weak abundance ratio) = 0,5.



Putting ourselves in the shoes of people on the slow road at $T = 52$, would they generally prefer to live in the post-labor stage of the fast track existing under our hypothesis? Given what we said earlier about the qualitative properties of both labor and non-labor activities, we think the answer would again be no, for traveling on the slow road to communism will bring a higher quality of life in both work and free time, regardless of the numbers for $L$ and $t$ obtaining in the above trade-off on our hypothesis.

This is somewhat of a paradox. It seems that when the transition to communism is underway for a long time, the labor and distributional criteria *M2* and *M4* both lose some of their importance in favour of the labor quality criterion *M3*. The paradox vanishes once the superior quality of non-labor activities on the slow road is considered as well. It then appears that the comprehensive qualitative improvement of activities in both the realms of necessity and freedom takes precedence, just as the ideal of communism ultimately requires (see 2.3). For this to happen, it must be the case that the amount of labor has become relatively small and the share of distribution according to need in total income has grown considerably over time, as is the case in our scenario, say, after $T = 40$. Moreover, for the realms of necessity and freedom to fuse into one at a high level of quality of life, high levels of real income will also be necessary. Since actual levels of real income in the slow version remain hidden in the SA model, we cannot be specific about the moment in time in which real income has risen sufficiently to ensure this, and it might therefore be the case that communism will be a reality somewhere after $T = 40$, but still before the post-labor stage at $T = 69$.

With this result, we are back to the material substrate of a communist society, which as Marx imagined, is a society 'rich in needs'. What ultimately counts for assessing the two versions of our scenario are the different compositions of discrete goods and services that support the



quality of life in each of them, or in Marxian parlance, support the 'enrichment of needs'. And as we argued in 2.2, these enriched needs are being formed by how technological innovations interact with the provision of consumption goods within markets and mechanisms of public financing. In terms of the 'richness of needs', then, the slow version of the second phase of the scenario, with its labor-friendly automated processes, is superior to the fast version and therefore the one preferred by the ideal of communism.

Our argument can be carried one final step further by asking how people living in the world of the slow version would assess their own situations at earlier and later moments in time. Would they really prefer to live in the post-labor stage at $T = 69$, or would they rather live in a world with an average working week of fourteen hours at $T = 40$? Many commentators have argued that even if labor becomes redundant in the sense that machines can perform better than humans across all tasks more cheaply, people will still want some tasks to remain in human hands and brains. Korinek and Juelfs (2022, 12) make this point forcefully: "… humans will always prefer to obtain certain services from other humans rather than from machines, for reasons that we may call 'nostalgic'. For example, humans may prefer not to replace the services of human priests, judges, or lawmakers." How long such reasons will persist is uncertain, and it is hard to say in advance what people several decades ahead in the future would think, if indeed labor had become fully redundant at that rather distant time. However, the 'nostalgic' demand for human services is not the only reason why people might want to keep on working, at possibly minute levels of monetary reward. For in the world of the slow road we are imagining here, the realms of freedom and of necessity will have fused into one in most respects. And what counts then is whether the quality of paid labor tasks is high enough for people with different levels of trained talent, in comparison with free-tine activities. If so, then the intrinsic value of engaging in some such tasks will be sufficient to



keep them wanting to stay in the 'realm of necessity'. This would conceivably give some, or even many, of those people reasons for preferring $T = 40$, or some other point before the post-labor stage on the slow road, to the final moment of full labor redundancy at $T = 69$.

To conclude, some scepticism about the relevance of numbers in Tables 3a and 3b is in order, especially the numbers running just up to the respective post-labor stages of both versions, say again after the turning point at $T = 40$, when both have $L = 30$. It may well be the case that at or somewhere after this point in time, the numbers of Table 3a for later periods on the slow road become misleading, if they are taken to suggest that the transition should necessarily go on until the post-labor stage, or beyond. This need not be so, as we have just argued. Even when people work fourteen hours a week on average, the quality of this work - when compared to that of free time - might be sufficiently high to think that communism has already been reached. In that case, so to speak, Keynes meets Marx. With respect to the fast track, there may be a different problem with Table 3b. For one thing, there is good reason to believe that the numbers for $G$ and $t$ recorded for the eleven years preceding the post-labor stage will be simply wrong, failing the normative motivation of policymakers on the fast track to set basic income at high ratios of per capita income for reasons of distributive justice. For another, the goal of elevating basic income in order to prevent involuntary underemployment in those eleven years will also have ceased to gain much attention of fast track-policymakers, since on the fast track, labor is on the way out anyhow and is expected to be fully redundant soon. Thus, while on the slow road of our scenario, the transition to communism might already be successful before the post-labor stage is reached, on the fast track it is likely to have already failed a while before reaching the - much earlier - post-labor stage.



## 6. Conclusions

In this concluding section we summarize and comment on our findings, trying to bring out what may be useful about the project of revisiting *Capitalist Road*, including its bearing on the immediate chances of installing a basic income, our third reason which was mentioned in the introduction and postponed until now. The main point of returning to the transition idea is that major aspects of it suddenly have taken a more concrete form in the digital age, and in particular following the AI revolution in technology. Many others are speculating on what might happen in the 'post-labor stage'. The value of our update, if any, is perhaps to make sense of what could be going on before paid labor has actually been eliminated and to present some concrete thoughts what this might mean in reality.

Of course, as we have noted, many economists staunchly disbelieve in a post-labor future. They may do so for good reasons. Our assumption in this paper is necessarily the one of Marx which underlies the transition idea: economic growth involves getting rid of necessary labor-time - so progress means shedding more old jobs than creating new ones. Therefore, we let the economy move to a post-labor stage simply because the subject matter of the transition to communism requires this to be so. But we also do hold a post-labor future to be a realistic possibility, while guessing that it might take at least half a century to get there. This may be just as well, for as argued in the last section, the fast track to a communist utopia can be easily derailed. Incidentally in sections 2 and 5, we have not attempted to recreate the entirety of Marx's utopian ideas on the state of unalienated men and women in the distant future of mankind in communist society, simply because the more lofty ones among these ideas go quite far beyond the terrain covered by *Capitalist Road*.[45]

---

[45] For example see Bertell Ollman (1976, 135) citing an early statement in the 1844 Manuscripts of how communism is the reunification of humans in their alienated states, and is defined as "…the positive transcendence of all estrangement - that is to say, the return of man from religion, family, state, etc., to his human, i.e. social mode of existence." See also Agnes Heller's (1974, 122, 125) characterizations of Marx's



The economic argument in *Capitalist Road* had no prominent role for automation and did not even explicitly include capital goods. Revisiting its arguments at present obviously requires an attempt to model automation, and we chose to adopt the clearcut distinction of the simple SA model between fixed capital which complements labor and automation capital which can fully substitute labor, while acknowledging the power of task-based models of automation in 3.2. The simulation in the baseline scenario of Table 1 in 4.1 is meant to illustrate how the model works. It has automation capital replacing labor at a very slow pace. Crude in many respects as the baseline scenario may be (see the caveats in 4.1), it has arguably credible initial values and parameters. And the model is flexible enough to run the other simulations of investment and technical change which enable us to focus on chances of reaching the transition to communism earlier, in our speculative scenario of section 5, a scenario which tries to work out what could happen when automation is speeding up, and when more billions than ever are invested in digital products and processes.

We compare relatively slow and fast simulations of technical change which can obtain after a period of disruption following a calm start in which, as yet, the economy is not labor-saving. This calm start is meant to show that when automation capital is introduced slowly, with skill-enhancing technologies, the displacement rate of labor can be equal or less than the rate at which new and profitable tasks are created. The point of inserting a 'period of disruption' in our scenario - it's essentially a black box containing unspecified interactions between

---

view on the 'irrelevance of equality and inequality as reciprocal determinations', and of the irrelevance of economic decision-making in the face of the communist harmony of interests: "In 'socialized man' the human species and the individual represent a realized unity. Every individual represents the species, and the species is represented in the human individual. The needs of 'socialized' human beings determine production - and this means that the human species itself makes the decisions." These final stages of communism go well beyond the relatively unambitious conceptions of utopia which Bostrom (2024, 126-7) characterizes as 'governance and culture utopias' and does not yet carry utopian states of society to the point of canceling the need for instrumental action in pursuit of daily goals and purposes. Bostrom's sophisticated account of different conceptions of utopia and his engaging fantasies about how to find meaning in life under radical abundance however, is highly useful for our purposes as well.



technical change and institutional arrangements - is to discuss two possible outcomes, leading to two 'capitalist roads'. In the first of these, technology is shaped to predominantly serve the interests of workers, following some market socialist reforms which we outline in our sketch of Textbox 2. This 'slow road' has labor benefitting from automation up to the point at which it ultimately becomes redundant. The second capitalist road essentially lies within welfare capitalism as we know it presently. It ends up putting automation predominantly in the service of capitalist interests. The hallmark of this 'fast track' is that it invests in automation processes and algorithms to make labor redundant sooner rather than later, and as a result of succeeding in this, it does not greatly benefit workers while labor is on its way out. The post-labor stages envisaged at the respective ends of these two imagined capitalist roads will of course reflect these differences.

Our main finding is hardly surprising. The slow road of the scenario is the most promising one - you need a decidedly labor-friendly and fairly reticent labor-replacing path of automation for transitioning to communism. In our discussion of this road, we indicate several ways in which the ideal of communism can be reached, or at least approximated, in the liberal-egalitarian interpretation of Marx set out in sec. 2. In particular we emphasise the value of progressively removing the 'separation' of free time outside the sphere of paid work and necessary time within that sphere. In our account of how the transition to communism unfolds along the slow road, this value moves to the forefront as labor gets phased out and as abundant material provisions satisfy increasingly richer needs more unconditionally. We also argue in 5.4 that with some luck, communism might be within reach even before the post-labor stage. This could be anytime between forty and seventy years from now on the scenario of the slow road.



Our attempt to sensibly update the original argument of *Capitalist Road*, with its resolute refusal to consider socialism as a precondition of communism, does come at a price. For even though our update may have become more convincing by playing the card of market socialism, its argument also seems to be less crisp, less provocatively fresh compared to the original argument. Do we then advocate turning the capitalist road to communism into a socialist one? To argue that 'socialism' - of whatever variety strong enough to usefully differentiate it from welfare capitalism - is needed from the very start to begin moving towards communism would of course be an unfortunate way of revisiting the original argument. Fortunately, this is not what needs to be argued. It is true that we think welfare capitalism in its current state of being under strong pressures of unequal outcomes and chances will not produce anything like communism over time, and indeed, this was our first reason for re-examining the argument of *Capitalist Road*. Therefore, inserting significant elements of socialism into the institutional structures of welfare capitalist societies along the way is important. We outlined such elements in section 5.1, and we think that in the coming years these elements can and must increasingly be realized in any case, partly because the time to stand up to excessive inequality is ripe, and partly because fast technological development is creating the need to adjust and align over a large arena of issues. But we concede that the degree of optimism needed to make our own somewhat more nuanced case for a possibly viable capitalist road to communism is hardly less than the one put forward by the original article. Our present update affirms in a different way that reaching communism does not require you to keep waiting for socialism. A prior transformation of welfare capitalism into some kind of readymade socialist regime, even a liberal-democratic market socialist one, is not necessary to start out on the road to communism. Our update also claims that progressive reforms are needed, but that as time goes on, the labels of 'capitalism' and 'socialism' are bound to become largely unimportant.



Finally, we return to our third reason for revisiting *Capitalist Road*. What does advocating basic income as the key policy that opens a long road to communism imply for the more immediate future? One way of addressing this is to recall our previous defence of replacing the traditional conditional arrangements of social security by unconditional transfers, set out in a friendly debate with Van Parijs and Vanderborght (van der Veen and Groot, 2019, 153-198; Van Parijs and Vandenborght, 2019, 247-261). That defence has a normative and a political-economic side. Starting with two normative desiderata: basic income must satisfy the principle of maximin opportunity, and be attractive in its own right according to a conception of the good which emphasizes the value of widening choice options and providing the right conditions for autonomous individual choices reflective of the social world.

Apart from a disagreement on whether basic income set at its highest sustainable level is the *only* way to achieve the maximin principle (thus ruling out as unjust conditional instruments of egalitarian distribution, such as wage subsidies), we share both normative requirements. We hold that, at present, unconditional basic income set at or above a decent level of subsistence can satisfy the two requirements far better than welfare granted on a work-conditional basis will ever be able to do. And we also applaud the way in which the second requirement has been argued by Van Parijs and Vanderborght (2019, Ch. 1) by explaining what the liberating social effects of a sufficiently generous basic income can be once it is durably implemented. In a nutshell, they explain how this enhances the bargaining power of individuals in paid work and family care, provides incentives to form communal living arrangements, and promotes flexibility in time allocation over a whole life. Indeed we believe that their vision of how unconditional transfers can in the immediate future serve a liberal conception of the good contains the 'seeds of communism'. In our present account of the key



role of basic income in bringing about a future communist society which embodies Marx's original ideal of the progressive suspension of the domains of paid work and free time, we extend that vision.

Our defence of basic income also contains a prudential element of political economy. For as noted in sec. 2 and 4.1, we also insist that basic income should be complemented by extensive social provisions of the normal sort within the welfare state, provisions which also must be placed on a more egalitarian basis, especially in education and vocational training. We placed such provisions in Dimension 4 in Textbox 2 above. Our reasons for stressing the need for them in debate with Van Parijs and Vanderborght are twofold. First, we claim that the beneficial effects of dispensing a fully unconditional basic income at subsistence level or above will be greatly increased when regular cash without strings attached is accompanied by these collective welfare benefits.

Secondly, claims derived from within social and Christian democratic ideology about the justice of equalizing opportunities for collective welfare - notably in health and education - compete with liberal claims calling for unconditional cash transfers. In a political conflict of priorities, the resource space for maximizing the level of these cash transfers is therefore somewhat limited. By forming political coalitions on the Left with good arguments for embedding basic income in an extensive welfare state which invoke the advantages of institutional complementarity, this conflict may be overcome, as we saw in 5.2. However, such arguments will be difficult to make as long as the level of unconditional income is below ruling subsistence norms, in need of being topped up by conditional supplementary transfers, and as long as there is a strong belief that raising the level of cash transfers and welfare state provisions together is economically unsustainable not only in the short run, but also in a



longer timeframe. To counter this belief is where our update of *Capitalist Road* may also become useful, and so we hope, politically inspiring.